\documentclass[11pt,onecolumn,journal]{IEEEtran}
\ifCLASSINFOpdf
   \usepackage[pdftex]{graphicx}
 \else
\fi
\usepackage{amsmath}
\usepackage{amssymb}
\usepackage{algorithm}
\usepackage{multirow}
\usepackage{array}
\usepackage{multirow,color}
\usepackage{epsfig}
\usepackage{epstopdf}
\begin{document}

\renewcommand\IEEEkeywordsname{Keywords}

\title{Study of Set-Membership Adaptive Kernel Algorithms }
\author{Andr{\'e} Flores and Rodrigo C. de Lamare

\thanks{Andr{\'e} Flores is with the Centre for Telecommunications Studies
(CETUC), PUC-Rio, Rio de Janeiro, Brazil and Rodrigo C. de Lamare is
with both CETUC and with the Department of Electronic Engineering,
University of York, UK. Part of this work has been presented at the
IEEE International Conference on Acoustics, Speech and Signal
Processing 2017. The emails of the authors are
andre.flores@cetuc.puc-rio.br and delamare@cetuc.puc-rio.br}}

\maketitle

\begin{abstract}
In the last decade, a considerable research effort has been devoted
to developing adaptive algorithms based on kernel functions. One of
the main features of these algorithms is that they form a family of
universal approximation techniques, solving problems with
nonlinearities elegantly. In this paper, we present data-selective
adaptive {kernel normalized least-mean square (KNLMS)} algorithms
that can increase their learning rate and reduce their computational
complexity. In fact, these methods deal with kernel expansions,
creating a growing structure also known as the dictionary, whose
size depends on the number of observations and their innovation.
{The algorithms described herein use an adaptive step-size to
accelerate the learning and can offer an excellent tradeoff between
convergence speed and steady state, which allows them to solve
nonlinear filtering and estimation problems with a large number of
parameters without requiring a large computational cost.} The
data-selective update scheme also limits the number of operations
performed and the size of the dictionary created by the kernel
expansion, saving computational resources and dealing with one of
the major problems of kernel adaptive algorithms. A statistical
analysis is carried out along with a computational complexity
analysis of the proposed algorithms.
 {Simulations show that the proposed KNLMS algorithms
outperform existing algorithms in examples of nonlinear system
identification and prediction of a time series originating from a
nonlinear difference equation.}
\end{abstract}



%
\IEEEpeerreviewmaketitle

\section{Introduction}

 {Adaptive filtering algorithms have been the focus of
a great deal of research in the past decades and the machine
learning community has embraced and further advanced the study of
these methods. In fact, adaptive algorithms are often considered
with linear structures, which limits their performance and does not
draw attention to nonlinear problems that can be solved in various
applications. In order to deal with nonlinear problems a family of
nonlinear adaptive algorithms based on kernels has been developed.
In particular, a kernel is a function that compares the similarity
between two inputs and can be used for filtering, estimation and
classification tasks. Kernel adaptive filtering (KAF) algorithms
have been tested in many different scenarios and applications
\cite{LiuPrincipeHaykin2010,Gil-Cacho2013,Gil-Cacho2012,Nakijama2012,RichardBermudezHoneine2009},
showing very good results. One of the main advantages of KAF
algorithms is that they are universal approximators
\cite{LiuPrincipeHaykin2010}, which gives them the ability to
address complex and nonlinear problems. However, their computational
complexity is much higher than their linear counterparts
\cite{LiuPrincipeHaykin2010}.}

One of the first KAF algorithms to appear, which is widely adopted
in the KAF family because of its simplicity, is the kernel
least-mean square (KLMS) algorithm proposed in
\cite{LiuPokharelPrincipe2008} and later extended in
\cite{Boboulis2011}. The KLMS algorithm has been inspired by the
least-mean square (LMS) algorithm and, thanks to its good
performance, led many researchers to work in the development of
kernel versions of conventional adaptive algorithms. For instance, a
kernel version of the NLMS algorithm has been proposed in
\cite{RichardBermudezHoneine2009} using a nonlinear regression
approach for time series prediction. In
\cite{LiuPrincipe2008,Slavakis2008}, the affine projection algorithm
(APA) has been used as the basis of the derivation of kernel affine
projection (KAP) algorithms. Adaptive projection algorithms using
kernel techniques have been reported in \cite{kapsg,Theodoridis11}.
The recursive least squares algorithm (RLS) has been extended in
\cite{Engel2004}, where the kernel recursive least squares (KRLS)
has been described. Later, the authors of \cite{Liu2009} proposed an
extension of the KRLS algorithm and the use of multiple kernels has
been studied in \cite{Pokharel2013} and \cite{Yukawa2012}.

 {Previously reported kernel algorithms have to deal
with kernel expansions, which increases significantly the
computational cost. In other words, they create a growing structure,
also called dictionary, where every new data input that arrives is
employed to compute the estimate of the desired output. The natural
problem that arises is that the time and computational cost required
to compute a certain output could exceed the tolerable limits for an
application.} Several criteria to manage the growing structure of
kernel algorithms have been proposed to solve this problem such as
algorithms with fixed dictionary size as studied in
\cite{VanVaerenberghViaSantamaria2006,VanVaerenberghSantamariaLiuEtAl2010}
and \cite{SheikholeslamiBerberidisGiannakis2015}. One of the most
simple criteria is the novelty criterion (NC), presented in
\cite{Platt1991}. Specifically, NC establishes two thresholds to
limit the size of the dictionary. Another method, the approximate
linear dependency (ALD) has been proposed in \cite{Engel2004} and
verifies if a new input can be expressed as a linear combination of
the elements stored before adding this input to the dictionary. The
coherence criterion (CC) has been described in
\cite{RichardBermudezHoneine2009} also to limit the size of the
dictionary based on the similarity of the inputs. A measure called
surprise criterion (SC) has been presented in \cite{Liu2009a} to
remove redundant data.

In this work, we present set-membership normalized kernel least-mean
square (SM-KNLMS) adaptive algorithms, which have been initially
reported in \cite{smknlms_2017,smknlms} and can provide a faster
learning than existing kernel-based algorithms and limit the size of
the dictionary without compromising performance. Unlike the
equivalent set-theoretic approach in \cite{Theodoridis11} the
set-membership algorithms presented here exploit variable step
sizes, which can lead to a faster learning performance. Similarly to
existing set-membership algorithms
\cite{FogelHuang1982,GollamudiNagarajKapoor1998,WernerDiniz2001,DinizWerner2003,Lamare2009,smce_conf,LamareDiniz2013,smce},
the proposed SM-KNLMS algorithms are equipped with variable step
sizes and perform sparse updates. We consider both centroid-based
SM-KNLMS (C-SM-KNLMS) and nonlinear regression-based SM-KNLMS
(NLR-KNLMS) algorithms, where the latter lends itself to statistical
analysis \cite{RichardBermudezHoneine2009}.  {Unlike existing
kernel-based adaptive algorithms the proposed SM-KNLMS algorithms
deal, in a natural way, with the kernel expansion because of the
data selectivity based on error bounds that they implement.} A
statistical analysis of the NLR-SM-KNLMS algorithm along with the
derivation of analytical formulas to predict the mean-square error
(MSE), and an analysis of their computational cost are carried out.
Simulations comparing the performance of the SM-KNLMS and existing
algorithms for several scenarios are then conducted.

In summary, the contributions of this work are:

\begin{itemize}

\item{The development of the proposed C-SM-KNLMS and NLR-SM-KNLMS algorithms.}

\item{A statistical analysis of the NLR-SM-KNLMS algorithm and the
development of analytical formulas to predict its performance.}

\item{A simulation study of the proposed C-SM-KNLMS, NLR-SM-KNLMS and existing algorithms for
several scenarios of interest.}
\end{itemize}

This paper is organized as follows. In Section II, the principles of
kernel methods and set-membership techniques are introduced. In
Section III, we review set-membership adaptive algorithms and
present the derivation of the proposed C-SM-KNLMS algorithm. Section
IV presents the proposed NLR-SM-KNLMS algorithm. Section V details
the statistical analysis of the NLR-SM-KNLMS algorithm and a
comparison of the computational complexity of the proposed and
existing algorithms. Section VI describes and discusses the
simulation results and Section VII contains the conclusions of this
work.

\section{Principles of Kernel Methods and Set-Membership Techniques}

Conventional adaptive algorithms work with linear structures,
limiting the performance that they can achieve and constraining the
number of problems that can be solved. Under this scope, a new
family of nonlinear adaptive algorithms based on kernels has been
developed \cite{LiuPrincipeHaykin2010}. The main objective of these
algorithms is to learn an arbitrary input-output mapping based on a
sequence of samples and a kernel. Basically, a kernel
$\kappa\left(\cdot,\cdot\right)$ is a function that measures the
similarity between two inputs and generally returns a real number.
Several kernel functions are described in the literature
\cite{LiuPrincipeHaykin2010}. Choosing a kernel function is
important because  {it} is equivalent to implicitly defining a
feature space where the algorithms are performed. Let us now
introduce two commonly used kernel functions. The first one is the
Gaussian kernel, defined by
\begin{equation}
\kappa\left(\mathbf{x},\mathbf{x}'\right)=\mbox{exp}\left(-\frac{\parallel\mathbf{x}-\mathbf{x}'\parallel^2}{2\nu^2}\right),
\end{equation}
where $\nu$ is the kernel bandwidth that specifies the shape of the
kernel function. Another important kernel function is the polynomial
kernel, given by
\begin{equation}
\kappa\left(\mathbf{x},\mathbf{x}'\right)=\left(\mathbf{x}^{\text{T}}\mathbf{x}'+1\right)^p,
\end{equation}
with $p\in\mathbb{N}$ known as the polynomial degree.

The relevant point about implementing kernel functions is that the
scalar product can be implicitly computed in the feature space by a
kernel evaluation, without explicitly using or even knowing the
mapping applied to the data \cite{MullerMikaRatschEtAl2001}.
 {This means that there is no need to perform any
operation on the high dimensional space, as long as the quantities
are expressed as an inner product. This approach is known as the
``kernel trick'' and allows us to compute scalar products in spaces,
where the computations are hard to perform.} As a result, we avoid a
significant increase of the computational complexity, which is one
of the major problems that arises when working with high dimensional
spaces. In particular we have
\begin{equation}
\kappa\left(\mathbf{x},\mathbf{x}'\right)=\left\langle\kappa\left(\cdot,\mathbf{x}\right),\kappa\left(\cdot,\mathbf{x}'\right)\right\rangle.
\end{equation}
To summarize, kernel adaptive algorithms map the data to a
high-dimensional space through kernels. Then, linear methods can be
applied on the transformed data to solve nonlinear problems.

Let us now consider an adaptive linear filtering problem with a sequence
of training samples given by $\left\{
\mathbf{x}\left(i\right),d\left(i\right)\right\} $, where
$\mathbf{x}\left(i\right)$ is the N-dimensional input vector and
$d\left(i\right)$ represents the desired response at time instant
$i$. The output of the adaptive linear filter is given by
\begin{equation}
y\left(i\right)=\mathbf{w}^{\text{T}}\left(i\right)\mathbf{x}\left(i\right),
\end{equation}
where $\mathbf{w}\left(i\right)$ is the weight vector with length
$N$.

We can extend linear models to nonlinear models by mapping the input
data into a high-dimensional space. In order to perform this
mapping, let us define a nonlinear transformation denoted by
$\varphi:\mathbb{R}^{N}\rightarrow \mathbb{F}$,
 {which maps the data in $\mathbb{R}^{N}$ to a
high-dimensional feature space $\mathbb{F}$ that performs the
nonlinear transformation}. Applying the transformation stated
before, we map both the input and the weights into the feature space
which results in
\begin{equation}
\boldsymbol{\varphi}\left(i\right)=\varphi\left(\mathbf{x}\left(i\right)\right),
\end{equation}
\begin{equation}
\boldsymbol{\omega}\left(i\right)=\varphi(\mathbf{w}\left(i\right)).
\end{equation}

We should emphasize that $\boldsymbol{\omega}\left(i\right)$ is now
a vector where each component is a function of the elements of
$\mathbf{w}\left(i\right)$, so that the dimension of
$\boldsymbol{\omega}\left(i\right)$ is greater than
$\mathbf{w}\left(i\right)$. The error generated by the system is
given by
\begin{equation}
e\left(i\right)=d\left(i\right)-
\boldsymbol{\omega}^{\text{T}}\left(i\right)
\boldsymbol{\varphi}\left(i\right).\label{eq:error}
\end{equation}

The main idea behind set-membership algorithms is to model a
function $\boldsymbol{\omega}\left(i\right)$, such that the
magnitude of the estimated error defined by \eqref{eq:error} is
upper bounded by a quantity $\gamma$. Assuming that the value of
$\gamma$ is appropriately chosen, there exist several functions that
satisfy the error requirement. In other words, any function leading
to an estimation error smaller than the defined threshold is an
adequate solution, resulting in a set of solutions. Otherwise if the
value of $\gamma$ is not properly chosen (if it is too small for
example), then there might be no solution.

Consider a set $\boldsymbol{\bar{S}}$ containing all the possible
input-desired signal pairs $\left\{
\boldsymbol{\varphi}\left(i\right),d\left(i\right)\right\} $ of
interest. Now we can define a set $\boldsymbol{\theta}$ with all the
possible functions leading to an estimation error bounded in
magnitude by $\gamma$. This set is known as the feasibility set and
is expressed by
\begin{equation}
\boldsymbol{\theta}=\bigcap_{\left\{ \boldsymbol{\varphi},d\right\}
\in\boldsymbol{\bar{S}}}\left\{
\boldsymbol{\omega}\in\mathbb{F}\:/\:|d-\boldsymbol{\omega}^{\text{T}}
\boldsymbol{\varphi}|\leq\gamma\right\}.
\end{equation}

Suppose now that we consider only the case in which only measured
data are available. Let us define a new set
$\mathcal{H}\left(i\right)$ with all the functions such that the
estimation error is upper bounded by $\gamma$ . The set is called
the constraint set and is mathematically defined by
\begin{equation}
\mathcal{H}\left(i\right)=\left\{ \boldsymbol{\omega}\in\mathbb{F}\:/\:|d\left(i\right)-\boldsymbol{\omega}^{\text{T}}\boldsymbol{\varphi}\left(i\right)|\leq\gamma\right\}. \label{eq:constraint set}
\end{equation}

It follows from \eqref{eq:constraint set} that, for each data pair,
there exists an associated constraint set. The set containing the
intersection of the constraint sets over all available time instants
is called exact membership set and is given by the following
equation:
\begin{equation}
\psi\left(i\right)=\bigcap_{k=0}^{i}\mathcal{H}\left(i\right).
\end{equation}
The exact membership set, $\psi\left(i\right)$, should become small
as the data containing new information arrives. This means that,
 {assuming stationary,} at some point the adaptive
algorithm will reach a state where
$\psi\left(i+1\right)=\psi\left(i\right)$, so that there is no need
to update $\boldsymbol{\omega}$. This happens because
$\psi\left(i\right)$ is already a subset of
$\mathcal{H}\left(i+1\right)$. As a result, the update of any
set-membership based algorithm is data dependent, saving resources,
a fact that is crucial in kernel-based adaptive algorithms because
of the growing structure that they create.


\section{Proposed Centroid-Based Set-Membership Kernel Normalized
Least-Mean-Square Algorithm}

In this section, we detail the derivation of the proposed C-SM-KNLMS
algorithm,  {which is motivated by the possibility of of saving
resources by not storing the zero coefficients in the parameter
vector.} In order to derive the C-SM-KLNMS algorithm, we check first
if the previous solution is outside the constraint set, i.e.,
$$|d\left(i\right)-\boldsymbol{\omega}^{\text{T}}\left(i\right)\boldsymbol{\varphi}\left(i\right)|>\gamma.$$
If the error exceeds the bound established, the algorithm performs
an update so that the \textit{a posteriori} estimated error lies in
$\mathcal{H}\left(i\right)$.

The derivation of the C-SM-KNLMS algorithm corresponds to solving
the following optimization problem \cite{altpow}:
\begin{equation}
\begin{split}
\min_{\boldsymbol{\omega}\left(i+1\right)} &
||\boldsymbol{\omega}\left(i+1\right)-\boldsymbol{\omega}\left(i\right)||^{2}
\\  {\rm subject ~~to}~~
& \boldsymbol{\omega}\left(i+1\right)\in\mathcal{H}\left(i\right),
\end{split}
\end{equation}
where the \textit{a posteriori} error $\xi_{p}\left(i\right)$ used
to build the constraint set $\mathcal{H}\left(i\right)$ is given by
\begin{equation}
\xi_{p}\left(i\right)=d\left(i\right)-\boldsymbol{\omega}^{\text{T}}\left(i+1\right)\boldsymbol{\varphi}\left(i\right)=\pm\gamma.\label{eq:posteriori error}
\end{equation}
As mentioned in \cite{LiuPrincipeHaykin2010}, the KNLMS update
equation is given by
\begin{equation}
\boldsymbol{\omega}\left(i+1\right)=\boldsymbol{\omega}\left(i\right)+\frac{\mu\left(i\right)}{\varepsilon+
||\boldsymbol{\varphi}\left(i\right)||^{2}}e\left(i\right)\boldsymbol{\varphi}\left(i\right),\label{eq:KNLMS}
\end{equation}
where $\mu\left(i\right)$ is the step-size that should be chosen to
satisfy the constraints and $\varepsilon$ is a small constant used
to avoid numerical problems. Substituting \eqref{eq:KNLMS} in
\eqref{eq:posteriori error} we arrive at:
\begin{equation}
\xi_{p}\left(i\right)=d\left(i\right) -
\boldsymbol{\omega}^{\text{T}}\left(i\right)\boldsymbol{\varphi}\left(i\right)
-\frac{\mu\left(i\right)}{\varepsilon+||\boldsymbol{\varphi}\left(i\right)||^{2}}
e\left(i\right)\boldsymbol{\varphi}^{\text{T}}\left(i\right)\boldsymbol{\varphi}\left(i\right)
\end{equation}
Using \eqref{eq:error} and replacing the dot products by kernel
evaluations, the previous equation turns into:
\begin{equation}
\xi_{p}\left(i\right)=e\left(i\right)-\mu\left(i\right)e
\left(i\right)\frac{\kappa\left(\mathbf{x}\left(i\right),
\mathbf{x}\left(i\right)\right)}{\varepsilon+\kappa\left(\mathbf{x}\left(i\right),
\mathbf{x}\left(i\right)\right)}=\pm\gamma.\label{eq:error
a posteriori SMKNLMS1}
\end{equation}
Assuming that the constant $\varepsilon$ is sufficiently small to
ensure that
\begin{equation}
\frac{\kappa\left(\mathbf{x}\left(i\right),
\mathbf{x}\left(i\right)\right)}{\varepsilon+\kappa\left(\mathbf{x}\left(i\right),\mathbf{x}\left(i\right)\right)}\approx1,
\end{equation}
then from  {Equation} \eqref{eq:error a posteriori SMKNLMS1}, we
have
\begin{equation}
\gamma=|e\left(i\right)\left(1-\mu\left(i\right)\right)|.
\end{equation}

If $\mu\left(i\right)$ takes values between 0 and 1, it follows
that:
\begin{equation}
|e\left(i\right)|\left(1-\mu\left(i\right)\right)=\gamma,
\end{equation}
\begin{equation}
\mu\left(i\right)=1-\frac{\gamma}{|e\left(i\right)|}.
\end{equation}
Taking into account that the update only occurs if the error is
greater than the specified bound then $\mu\left(i\right)$ is
described by
\begin{equation}
\mu\left(i\right)=\begin{cases}
\begin{array}{c}
1-\frac{\gamma}{|e\left(i\right)|}\\
0
\end{array} & \begin{array}{c}
|e\left(i\right)|>\gamma,\\
\mbox{otherwise.}
\end{array}\end{cases}\label{eq:step size}
\end{equation}
We can then compute $\boldsymbol{\omega}$ recursively
as follows:
\begin{align}
\boldsymbol{\omega}\left(i+1\right)  =&  \boldsymbol{\omega}\left(i-1\right)+\frac{\mu\left(i-1\right)e\left(i-1\right)}{\varepsilon+||\boldsymbol{\varphi}\left(i-1\right)||^{2}}\boldsymbol{\varphi}\left(i-1\right)\nonumber\\&+\frac{\mu\left(i\right)e\left(i\right)}{\varepsilon+||\boldsymbol{\varphi}\left(i\right)||^{2}}\boldsymbol{\varphi}\left(i\right)\nonumber \\
 & \vdots\nonumber \\
 \boldsymbol{\omega}\left(i+1\right) =& \boldsymbol{\omega}\left(0\right)+\sum_{k=1}^{i}\frac{\mu\left(k\right)}{\varepsilon+||\boldsymbol{\varphi}\left(k\right)||^{2}}e\left(k\right)\boldsymbol{\varphi}\left(k\right)
\end{align}
Setting $\boldsymbol{\omega}\left(0\right)$ to zero leads to:
\begin{equation}
\boldsymbol{\omega}\left(i+1\right)=\sum_{k=1}^{i}\frac{\mu\left(k\right)}{\varepsilon+||\boldsymbol{\varphi}\left(k\right)||^{2}}e\left(k\right)\boldsymbol{\varphi}\left(k\right).
\end{equation}
The output
$f\left(\boldsymbol{\varphi}\left(i+1\right)\right)=\boldsymbol{\omega^{\text{T}}}\left(i+1\right)\boldsymbol{\varphi}\left(i+1\right)$
of the filter to a new input $\boldsymbol{\varphi}\left(i+1\right)$
can be computed as:
\begin{align}
f\left(\boldsymbol{\varphi}\left(i+1\right)\right) & =  \left[\sum_{k=1}^{i}\frac{\mu\left(k\right)}{\varepsilon+||\boldsymbol{\varphi}\left(k\right)||^{2}}e\left(k\right)\boldsymbol{\varphi}^{\text{T}}\left(k\right)\right]\boldsymbol{\varphi}\left(i+1\right),\nonumber\\
 & =  \sum_{k=1}^{i}\frac{\mu\left(k\right)}{\varepsilon+||\boldsymbol{\varphi}\left(k\right)||^{2}}e\left(k\right)\boldsymbol{\varphi}^{\text{T}}\left(k\right)\boldsymbol{\varphi}\left(i+1\right).
\end{align}
Using the kernel trick we obtain
\begin{equation}
f\left(\boldsymbol{\varphi}\left(i+1\right)\right)=\sum_{k=1}^{i}\frac{\mu\left(k\right)e\left(k\right)}{\varepsilon+\kappa\left(\mathbf{x}\left(k\right),\mathbf{x}\left(k\right)\right)}\kappa\left(\mathbf{x}\left(k\right),\mathbf{x}\left(i+1\right)\right),\label{eq:update equation}
\end{equation}
where $\mu\left(k\right)$ is given by \eqref{eq:step size}. Let us
define a coefficient vector $\boldsymbol{a}\left(i\right)$ to store
in each of its elements the following product:
\begin{equation}
\left[\boldsymbol{a}\left(i\right)\right]_k=\mu\left(k\right)e\left(k\right),\label{eq:coeficientes}
\end{equation}
so that \eqref{eq:update equation} becomes:
\begin{equation}
f\left(\boldsymbol{\varphi}\left(i+1\right)\right)=\sum_{k=1}^{i}\frac{a_k\left(i\right)}{\varepsilon+\kappa\left(\mathbf{x}\left(k\right),\mathbf{x}\left(k\right)\right)}\kappa\left(\mathbf{x}\left(k\right)\boldsymbol{,}\mathbf{x}\left(i+1\right)\right).\label{eq:SM-KNLMS}
\end{equation}
Eqs. \eqref{eq:error}, \eqref{eq:step size},\eqref{eq:coeficientes},
and \eqref{eq:SM-KNLMS} summarize the proposed C-SM-KNLMS algorithm.
We set the initial values of $\boldsymbol{a}$ to zero. As new inputs
arrive we can calculate the output of the system with
\eqref{eq:SM-KNLMS}. Then the error may be computed with
\eqref{eq:error} and if it exceeds the bound we compute the
step-size with \eqref{eq:step size}. The vector
$\boldsymbol{a}\left(i\right)$ are updated with
\eqref{eq:coeficientes}. Note that some coefficients may be zero due
to the data selectivity of C-SM-KNLMS. We do not need to store the
zero coefficients as they do not contribute to the output, resulting
in saving of resources. This means that the dictionary at time
instant $i$, denoted by $\mathcal{\boldsymbol{C}}\left(i\right)$,
has only $m$ elements, with $m<i$. Each column of the dictionary,
denoted by $\mathcal{\boldsymbol{c}}_{j}$, contains the input that
is used in the $k$th update. We can now rewrite \eqref{eq:SM-KNLMS}
as follows:
\begin{equation}
\boldsymbol{\omega^{\text{T}}}\left(i+1\right)\boldsymbol{\varphi}\left(i+1\right)=\sum_{k=1}^{m}\frac{a_{k}\left(i\right)}{\varepsilon+\kappa\left(\mathcal{\boldsymbol{c}}_{k},\mathcal{\boldsymbol{c}}_{k}\right)}\kappa\left(\mathbf{x}\left(i\right)\boldsymbol{,}\mathcal{\boldsymbol{c}}_{k}\right)
\end{equation}
This is an important result because it controls the growing network
created by the algorithm
\cite{intadap,jio,spa,jidf,sjidf,mfsic,armo,tds,baplnc,l1stap,L1reg,mmimo,barc,mbdf,wence,rrstap,mbthp,ccg,wcccm,ccmmwf,vfap,locsme,dce,lbrcodec,bfpeg,saalt,wlbeam,bfidd,rdrcb,did,saabf,als,wlbd,memd,rrmser,rrecho,doajio,okspme,damdc,rrdoa,kaesprit}.
In stationary environments the algorithm will limit the growing
structure. Algorithm \ref{alg:Set-Membership-Normalized-Kernel}
summarizes the proposed C-SM-KNLMS algorithm.  {In particular, the
computational complexity of C-SM-KNLMS grows over time with the
increase of $m$, as illustrated by step 7 in Algorithm
\ref{alg:Set-Membership-Normalized-Kernel}. However, we also note
that the standard KNLMS algorithm exhibits such behavior with
regards to the computational complexity. Unlike the standard KNLMS
the proposed C-SM-KNLMS algorithm only performs an update when there
is innovation according to the error bound, which limits the
increment of $m$ and consequently the increase in computational
complexity.}

\begin{algorithm}[H]
\caption{Proposed C-SM-KNLMS algorithm
\label{alg:Set-Membership-Normalized-Kernel}}
\textbf{Initialization}

1.Choose $\gamma$ , $\varepsilon$ and $\kappa$.

2.$\mathcal{\boldsymbol{C}}\left(1\right)=\left\{ \mathbf{x}\left(1\right)\right\} $

3.$\mu\left(1\right)=1-\frac{\gamma}{|d\left(1\right)|}$

4.$a_{1}\left(1\right)=\mu\left(1\right)d\left(1\right)$

5.$m=1$

\textbf{Computation }

6.\textbf{while} $\left\{ \mathbf{x}\left(i\right),d\left(i\right)\right\} $
available \textbf{do}:

~~~~~~~~~~~~\%Compute the output

7.~~~~~~~~~$f_{i-1}(\mathbf{x}\left(i\right))=\sum_{k=1}^{m}\frac{a_{k}\left(i\right)}{\varepsilon+\kappa\left(\mathcal{\boldsymbol{c}}_{k},\mathcal{\boldsymbol{c}}_{k}\right)}\kappa\left(\mathbf{x}\left(i\right)\boldsymbol{,}\mathcal{\boldsymbol{c}}_{k}\right)$

~~~~~~~~~~~~\%Compute the error

8.~~~~~~~~~$e\left(i\right)=d\left(i\right)-f_{i-1}(\mathbf{x}\left(i\right))$

9.~~~~~~~~~\textbf{if} $|e\left(i\right)|>\gamma$

~~~~~~~~~~~~~~~~~~~\%Compute the step-size

10.~~~~~~~~~~~~~~~$\mu\left(i\right)=1-\frac{\gamma}{|e\left(i\right)|}$

~~~~~~~~~~~~~~~~~~~\%Update the coefficients

11.~~~~~~~~~~~~~~~$\boldsymbol{a}\left(i+1\right)=\left[\begin{array}{c}
\boldsymbol{a}\left(i\right)\\
0
\end{array}\right]+\left[\begin{array}{c}
\boldsymbol{0}\\
\mu\left(i\right)e\left(i\right)
\end{array}\right]$

~~~~~~~~~~~~~~~~~~~\%Store the new center

12.~~~~~~~~~~~~~~~$\boldsymbol{\mathcal{C}}\left(i+1\right)=\left\{ \boldsymbol{\mathcal{C}}\left(i\right),\mathbf{x}\left(i\right)\right\} $

13.~~~~~~~~~~~~~~~$m=m+1$

14.~~~~~~~\textbf{else}

15.~~~~~~~~~~~~~~~$\mu\left(i\right)=0$

16.~~~~~~~~~~~~~~~$\boldsymbol{a}\left(i+1\right)=\boldsymbol{a}\left(i\right)$

17.~~~~~~~~~~~~~~~$\boldsymbol{\mathcal{C}}\left(i+1\right)=\boldsymbol{\mathcal{C}}\left(i\right)$

18.~~~~~~~\textbf{end if}

19.\textbf{end while}
\end{algorithm}

\section{Proposed Nonlinear Regression-Based SM-KNLMS Algorithm}

In this section, we follow a nonlinear regression approach as
described in
\cite{RichardBermudezHoneine2009,CoelhoNascimentoQueirozEtAl2015},
to develop an alternative SM-KNLMS algorithm, denoted NLR-SM-KNLMS
algorithm.

Let us define a function $\psi\left(\cdotp\right)$ on a feature space 
which, given an input vector $\mathbf{x}\left(i\right)$ generates the model
output $\psi\left(\mathbf{x}\left(i\right)\right)$. Our problem
is now reduced to finding the function $\psi\left(\cdotp\right)$
that minimizes the sum of the square error between the desired response
and the model output as described by
\begin{equation}
\min_{\psi\in\mathcal{H}}\sum_{k=1}^{i}|d\left(k\right)-\psi\left(\mathbf{x}\left(k\right)\right)|^{2}
\end{equation}
The representer theorem \cite{ScholkopfHerbrichSmola2001} states
that the function $\psi\left(\cdotp\right)$ can be expressed as a
kernel expansion which depends on the available data, so that:
\begin{equation}
\psi\left(\cdotp\right)=\sum_{k=1}^{i}a_{k}\kappa\left(\cdotp,\mathbf{x}\left(k\right)\right).
\end{equation}
In order to derive the NLR-SM-KNLMS algorithm we need to solve the
following optimization problem:
\begin{equation}
\min_{\boldsymbol{a}}\parallel\boldsymbol{d}-\boldsymbol{Ka}\parallel^{2},
\end{equation}
\noindent where  {${\boldsymbol a} \in \mathbb{R}^{m}$ is the
parameter vector to be computed, ${\boldsymbol d} \in
\mathbb{R}^{m}$ is the vector with the desired signal and
$\boldsymbol{K} \in \mathbb{R}^{m \times m}$} is the Gram matrix
containing at each row $i$ and each column $j$ the kernel
evaluations of the input data denoted by $\kappa_{ij}$, where we
have
\begin{equation}
\left[\boldsymbol{K}\right]_{ij}=\kappa_{ij}=\kappa\left(\mathbf{x}\left(i\right),\mathbf{x}\left(j\right)\right).
\end{equation}
Let us now  consider the case where we have a dictionary of size $m$
so that $\boldsymbol{K}\in\mathbb{R}^{m\times m}$. Consider also a
vector $\boldsymbol{\kappa_{\delta}}\left(i\right)$ that contains
the kernel evaluations between the input data at time $i$ and every
input stored in the dictionary at time $i>m$ with
$\mathcal{\boldsymbol{c}}_{j}\neq\mathbf{x}\left(i\right)$ for
$j=1,\cdots,m$, given by
\begin{equation}
\boldsymbol{\kappa_{\delta}}\left(i\right)=\left[\begin{array}{c}
\kappa\left(\mathbf{x}\left(i\right),\mathcal{\boldsymbol{c}}_{1}\right)\\
\kappa\left(\mathbf{x}\left(i\right),\mathcal{\boldsymbol{c}}_{2}\right)\\
\vdots\\
\kappa\left(\mathbf{x}\left(i\right),\mathcal{\boldsymbol{c}}_{m+1}\right)
\end{array}\right],\label{eq:kernel vector dictionary and actual input}
\end{equation}
 {where $\boldsymbol{\kappa_{\delta}}\left(i\right)$
is used in the computation of an inner product with ${\boldsymbol
a}(i+1) \in \mathbb{R}^{m+1}$.} Using the minimum norm approach to
obtain the NLR-SM-KNLMS algorithm, the constrained optimization
problem becomes:
\begin{gather}
\min_{\boldsymbol{a}}\parallel\boldsymbol{a}\left(i+1\right)-\boldsymbol{a}\left(i\right)\parallel^{2}\nonumber\\
\mbox{subject to}\nonumber\\
\mid d\left(i\right)-\boldsymbol{\kappa_{\delta}}^{\text{T}}\left(i\right)\mathbf{\boldsymbol{a}}\left(i+1\right)\mid=0.
\end{gather}
Using the method of Lagrange multipliers, we have
\begin{equation}
\mathcal{L}(\boldsymbol{a},\boldsymbol{\lambda})=\parallel\boldsymbol{{a}}\left(i+1\right)-\boldsymbol{a}\left(i\right)\parallel^{2}+\lambda\left(d\left(i\right)-\boldsymbol{\kappa_{\delta}}^{\text{T}}\left(i\right){\boldsymbol{a}}\left(i+1\right)\right).
\end{equation}
Calculating the gradient with respect to to $\boldsymbol{a}\left(i+1)\right)$ and $\lambda$, we obtain
\begin{align}
\frac{\partial\mathcal{L}(\boldsymbol{a},\boldsymbol{\lambda})}{\partial\boldsymbol{a}\left(i+1\right)}&=\left(\boldsymbol{{a}}\left(i+1\right)-\boldsymbol{a}\left(i\right)\right)-\lambda\boldsymbol{\kappa_{\delta}}\left(i\right)=\mathbf{0},\label{eq:derivada de alfa KNLMS}\\
\frac{\partial\mathcal{L}(\boldsymbol{a},\boldsymbol{\lambda})}{\partial\lambda}&=d\left(i\right)-\boldsymbol{\kappa_{\delta}}^{\text{T}}\left(i\right){\boldsymbol{a}}\left(i+1\right)=0.\label{eq:multiplicador de lagrange KNLMS}
\end{align}
From equation \eqref{eq:derivada de alfa KNLMS} we obtain:
\begin{equation}
\lambda\boldsymbol{\kappa_{\delta}}\left(i\right)=\left(\boldsymbol{{a}}\left(i+1\right)-\boldsymbol{a}\left(i\right)\right),
\end{equation}
\begin{equation}
\lambda\boldsymbol{\kappa_{\delta}}^{\text{T}}\left(i\right)\boldsymbol{\kappa_{\delta}}\left(i\right)=\boldsymbol{\kappa_{\delta}}^{\text{T}}\left(i\right)\left(\boldsymbol{{a}}\left(i+1\right)-\boldsymbol{a}\left(i\right)\right).
\end{equation}
Substituting \eqref{eq:multiplicador de lagrange KNLMS} in the
equation above we get:
\begin{equation}
\lambda\parallel\boldsymbol{\kappa_{\delta}}\left(i\right)\parallel^{2}=\left(d\left(i\right)-\boldsymbol{\kappa_{\delta}}^{\text{T}}\left(i\right)\boldsymbol{a}\left(i\right)\right),
\end{equation}
\begin{equation}
\lambda=\frac{1}{\parallel\boldsymbol{\kappa_{\delta}}\left(i\right)\parallel^{2}}\left(d\left(i\right)-\boldsymbol{\kappa_{\delta}}^{\text{T}}\left(i\right)\boldsymbol{a}\left(i\right)\right).
\end{equation}
Finally, replacing $\lambda$ in equation \eqref{eq:derivada de alfa
KNLMS} we obtain the NLR-SM-KNLMS update recursion for the
coefficients, which is expressed as follows:
\begin{equation}
\boldsymbol{a}\left(i+1\right)=\boldsymbol{a}\left(i\right)+\frac{1}{\parallel\boldsymbol{\kappa_{\delta}}\left(i\right)\parallel^{2}}\left(d\left(i\right)-\boldsymbol{\kappa_{\delta}}^{\text{T}}\left(i\right)\boldsymbol{a}\left(i\right)\right)\boldsymbol{\kappa_{\delta}}\left(i\right).
\end{equation}
When using the NLR-SM-KNLMS algorithm, the update only occurs when
the error represented by
$d\left(i\right)-\boldsymbol{\kappa_{\delta}}\left(i\right)^{\text{T}}\boldsymbol{a}\left(i\right)$
exceeds the threshold $\gamma$. In this case, the dictionary size
should be increased by one as well as the length of the vector
$\boldsymbol{a}$. The update recursion is given by
\begin{equation}
\boldsymbol{a}\left(i+1\right)=\left[\begin{array}{c}
\boldsymbol{a}\left(i\right)\\
0
\end{array}\right]+\frac{\mu\left(i\right)}{\varepsilon+\parallel\boldsymbol{\kappa_{\delta}}\left(i\right)\parallel^{2}}e\left(i\right)
\boldsymbol{\kappa_{\delta}}\left(i\right), \label{eq:update non
linear regresion smknlms1}
\end{equation}
where
$e\left(i\right)=d\left(i\right)-\boldsymbol{\kappa_{\delta}}^{\text{T}}\left(i\right)\left[\begin{array}{c}
\boldsymbol{a}\left(i\right)\\
0
\end{array}\right].$

Let us now define the \textit{a posteriori} error as follows:
\begin{equation}
\xi_{p}\left(i\right)=d\left(i\right)-\boldsymbol{\kappa_{\delta}}^{\text{T}}\left(i\right)\boldsymbol{a}\left(i+1\right)=\pm\gamma.
\end{equation}
Substituting equation \eqref{eq:update non linear regresion
smknlms1} in the last equation and assuming that
$\frac{\parallel\boldsymbol{\kappa_{\delta}}\left(i\right)\parallel^{2}}{\varepsilon+\parallel\boldsymbol{\kappa_{\delta}}\left(i\right)\parallel^{2}}\approx1$,
we have
\begin{equation}
d\left(i\right)-\boldsymbol{\kappa_{\delta}}^{\text{T}}\left(i\right)\left[\begin{array}{c}
\boldsymbol{a}\left(i\right)\\
0
\end{array}\right]-\mu\left(i\right)e\left(i\right)=\pm\gamma.
\end{equation}
Simplifying the terms, we obtain \vspace{-3mm}
\begin{align}
\gamma=&e\left(i\right)-\mu\left(i\right)e\left(i\right),\nonumber\\
=&e\left(i\right)\left(1-\mu\left(i\right)\right).
\end{align}
From the last equation we obtain an expression for the step-size, which is given by
\begin{equation}
\mu\left(i\right)=\begin{cases}
\begin{array}{c}
1-\frac{\gamma}{\mid e\left(i\right)\mid}\\
0
\end{array} & \begin{array}{c}
\mid e\left(i\right)\mid>\gamma,\\
\mbox{otherwise}.
\end{array}\end{cases}
\end{equation}
If the error does not exceed the threshold $\gamma$, the size of the
dictionary remains the same and no coefficients update is performed,
only the output of the model is calculated for the new input. The
pseudo-code for the NLR-SM-KNLMS algorithm is shown in Algorithm
\ref{alg:SM-KNLMS non linear regression}.

%
%

\begin{algorithm}[H]
\caption{Nonlinear Regression SM-KNLMS Algorithm}\label{alg:SM-KNLMS non linear regression}

\textbf{Initialization}

1.Choose $\gamma$ , $\varepsilon$ and $\kappa$.


2.$\mu\left(1\right)=1-\frac{\gamma}{|d\left(1\right)|}$

3.$\boldsymbol{a}\left(1\right)=0$

4.$m=1$

5.$\boldsymbol{\kappa_{\delta}}\left(1\right)=\kappa\left(\mathbf{x}\left(1\right),\mathbf{x}\left(1\right)\right)$

\textbf{Computation }

6.\textbf{while} $\left\{ \mathbf{x}\left(i\right),d\left(i\right)\right\} $
available \textbf{do}:

~~~~~\%Compute vector $\boldsymbol{\kappa_{\delta}}\left(i\right)$

7.~~~$\boldsymbol{\kappa_{\delta}}\left(i\right)=\left\{ \kappa\left(\mathbf{x}\left(i\right),\mathbf{x}\left(\delta_{1}\right)\right),\ldots,\kappa\left(\mathbf{x}\left(i\right),\mathbf{x}\left(\delta_{m}\right)\right)\right\} $

~~~~~~\%Compute the output

8.~~~$y\left(i\right)=\boldsymbol{\kappa_{\delta}}^{\text{T}}\left(i\right)\boldsymbol{a}\left(i\right)$

~~~~~\%Compute the error

9.~~~$e\left(i\right)=d\left(i\right)-y\left(i\right)$

10.~~\textbf{if} $|e\left(i\right)|>\gamma$

~~~~~~~~~~\%Store the new center

11.~~~~~~~$\boldsymbol{\kappa_{\delta}}\left(i\right)=\left\{ \kappa\left(\mathbf{x}\left(i\right),\mathbf{x}\left(\delta_{1}\right)\right),\ldots,\kappa\left(\mathbf{x}\left(i\right),\mathbf{x}\left(\delta_{m+1}\right)\right)\right\} $

~~~~~~~~~~\%Store the step-size

12.~~~~~~~$\mu\left(i\right)=1-\frac{\gamma}{|e\left(i\right)|}$

~~~~~~~~~~\%Update the coefficients

13.~~~~~~~$\boldsymbol{a}\left(i+1\right)=\left[\begin{array}{c}
\boldsymbol{a}\left(i\right)\\
0
\end{array}\right]+\frac{\mu\left(i\right)}{\varepsilon+\parallel\boldsymbol{\kappa_{\delta}}\left(i\right)\parallel^{2}}e\left(i\right)\boldsymbol{\kappa_{\delta}}\left(i\right)$

14.~~~~~~~$m=m+1$

15.~~~\textbf{else}

16.~~~~~~~$\mu\left(i\right)=0$

17.~~~~~~~$\boldsymbol{a}\left(i+1\right)=\boldsymbol{a}\left(i\right)$

18.~~~\textbf{end if}

19.\textbf{end while}
\end{algorithm}

\section{Analysis}
\label{sec:Statistical Analysis}

In this section, we consider a statistical analysis of the
NLR-SM-KNLMS algorithm along with a computational complexity
comparison among the proposed and existing algorithms.

\subsection{Computational complexity}

The computational complexity of the proposed algorithms and the KLMS
algorithm is detailed in Table I. We consider real-valued data and
the cost is given in terms of the number of multiplications and
additions per iteration as a function of $N$, $m$ and the update
rate (UR). Moreover, the algorithms use a maximum fixed size for the
dictionary, which means that the computational complexity only
varies before reaching steady-state.

\begin{table}[!h]
\label{computable}
\begin{centering}\caption{Computational Cost per update Iteration}
\begin{tabular}{|c|c|c|}
\hline \textbf{Algorithm} & \textbf{Additions (+)} &
\textbf{Multiplications (x)}\tabularnewline \hline \hline KLMS & $
{m(N+1)+1}$ & $ {m(N+1)}$ \tabularnewline \hline KNLMS (Regression)
& $ {m(2N+1)+2}$ & $ {m(2N+1)+1}$ \tabularnewline \hline
 {C-SM-KNLMS (Algorithm 1)} & $ {m(2N)+1
+UR(1)}$ & $ {m(2N+1)+UR(1)}$\tabularnewline \hline
 {NLR-SM-KNLMS (Algorithm 2)} &
$ {(m+1)(N-1)+1+UR(N+2m+1)}$ & $
{(m+1)(2N)+UR(N+m+2)}$\tabularnewline \hline
\end{tabular}
\par
\end{centering}
\end{table}

\subsection{Statistical Analysis}

In this section, we consider a statistical analysis of the NLR-KNLMS
algorithm with a Gaussian kernel in a stationary environment, which
means that $\varphi\left(\mathbf{x}\left(i\right)\right)$ is
stationary for $\mathbf{x}\left(i\right)$ stationary
\cite{ChenGaoRichardEtAl2014}. We focus on the analysis of the
NLR-SM-KNLMS algorithm rather than C-SM-KNLMS because the former
lends itself to statistical analysis, as explained in
\cite{ChenGaoRichardEtAl2014}.

Several nonlinear systems used to model practical situations, such
as Wiener and Hammerstein systems, satisfy this assumption. The
system inputs are N-dimensional, independent and identically
distributed Gaussian vectors $\mathbf{x}\left(i\right)$ with
zero-mean and variance equal to $\sigma_x^2$. Let us denote the
autocorrelation matrix of the input vectors by
$\mathbf{R}_{xx}=\mathbb{E}\left[\mathbf{x}\left(i\right)\mathbf{x}^{\text{T}}\left(i\right)\right]$,
so that
$\mathbb{E}\left[\mathbf{x}\left(i-k\right)\mathbf{x}^{\text{T}}\left(i-l\right)\right]=\mathbf{0}$
for $k\neq l$. However the components of the input vector can be
correlated. Let us also consider a dictionary of fixed size M and
the vector $\boldsymbol{\kappa_{\delta}}\left(i\right)$ previously
defined in equation \eqref{eq:kernel vector dictionary and actual
input}. We assume that the vectors constituting the dictionary may
change at each iteration following some dictionary updating scheme.
The vectors composing the dictionary are statistically independent
because $\mathbf{x}\left(\delta_{j}\right)\neq\mathbf{x}\left(\delta_{k}\right)$
for $j\neq k$.

The estimated output of the system is described by
\begin{equation}
y\left(i\right)=\boldsymbol{a}^{\text{T}}\left(i\right)\boldsymbol{\kappa_{\delta}}\left(i\right).
\end{equation}

The corresponding estimation error is given by
\begin{equation}
e\left(i\right)=d\left(i\right)-y\left(i\right).
\end{equation}
Squaring the equation above and taking the expected value results
in the MSE:
\begin{align}
J_{\rm ms}\left(i\right) = & \mathbb{E}\left[e^{2}\left(i\right)\right]\nonumber\\
  = & \mathbb{E}\left[d^{2}\left(i\right)\right]-2\mathbf{p}_{kd}^{\text{T}}\boldsymbol{a}\left(i\right)+\boldsymbol{a}^{\text{T}}\left(i\right)\mathbf{R}_{kk}\boldsymbol{a}\left(i\right),
\end{align}
where $\mathbf{R}_{kk}=\mathbb{E}\left[\boldsymbol{\kappa_{\delta}}\left(i\right)\boldsymbol{\kappa_{\delta}}^{\text{T}}\left(i\right)\right]$
represents the correlation matrix of the kernelized input, and $\mathbf{p}_{kd}=\mathbb{E}\left[d\left(i\right)\boldsymbol{\kappa_{\delta}}\left(i\right)\right]$
is the cross-correlation vector between $\boldsymbol{\kappa_{\delta}}\left(i\right)$
and $d\left(i\right)$. In \cite{ParreiraBermudezRichardEtAl2012,ParreiraBermudezRichardEtAl2011}
it is shown that $\mathbf{R}_{kk}$ is positive definite. Thus, the
Wiener solution and the minimum MSE are obtained as follows:
\begin{eqnarray}
\boldsymbol{a}_{o} & = & \mathbf{R}_{kk}^{-1}\mathbf{p}_{kd}\\
J_{\rm min} & = &
\mathbb{E}\left[d^{2}\left(i\right)\right]-\mathbf{p}_{kd}^{\text{T}}\mathbf{R}_{kk}^{-1}\mathbf{p}_{kd},
\end{eqnarray}
The entries of the correlation matrix $\mathbf{R}_{kk}$ are given by

\begin{equation}
\left[\mathbf{R}_{kk}\right]_{jl}=\begin{cases}
\mathbb{E}\left[\kappa^{2}\left(\mathbf{x}\left(i\right),\mathbf{x}\left(\delta_{j}\right)\right)\right] & j=l\\
\mathbb{E}\left[\kappa\left(\mathbf{x}\left(i\right),\mathbf{x}\left(\delta_{j}\right)\right)\kappa\left(\mathbf{x}\left(i\right),\mathbf{x}\left(\delta_{l}\right)\right)\right] & j\neq l
\end{cases}
\end{equation}

Let us define the following products:

\begin{equation}
\left\Vert \mathbf{x}\left(i\right)-\mathbf{x}\left(\delta_{j}\right)\right\Vert ^{2}=\boldsymbol{y}_{2}^{\text{T}}\mathbf{Q}_{2}\boldsymbol{y}_{2}
\end{equation}
\begin{equation}
\left\Vert \mathbf{x}\left(i\right)-\mathbf{x}\left(\delta_{j}\right)\right\Vert ^{2}+\left\Vert \mathbf{x}\left(i\right)-\mathbf{x}\left(\delta_{l}\right)\right\Vert ^{2}=\boldsymbol{y}_{3}^{\text{T}}\mathbf{Q}_{3}\boldsymbol{y}_{3},
\end{equation}
where
\begin{equation}
\boldsymbol{y}_{2}=\left[\begin{array}{cc}
\mathbf{x}^{\text{T}}\left(i\right) & \mathbf{x}^{\text{T}}\left(\delta_{j}\right)\end{array}\right]^{\text{T}},
\end{equation}
\begin{equation}
\boldsymbol{y}_{3}=\left[\begin{array}{ccc}
\mathbf{x}^{\text{T}}\left(i\right) & \mathbf{x}^{\text{T}}\left(\delta_{j}\right) & \mathbf{x}^{\text{T}}\left(\delta_{l}\right)\end{array}\right]^{\text{T}},
\end{equation}
\begin{equation}
\mathbf{Q}_{2}=\left[\begin{array}{cc}
\thinspace\mathbf{\thinspace\thinspace\thinspace I} & -\mathbf{I}\\
-\mathbf{I} & \thinspace\mathbf{\thinspace\thinspace\thinspace I}
\end{array}\right],
\end{equation}
\begin{equation}
\mathbf{Q}_{3}=\left[\begin{array}{ccc}
\thinspace2\mathbf{I} & -\mathbf{I} & -\mathbf{I}\\
-\mathbf{I} & \mathbf{\thinspace\thinspace\thinspace\thinspace I} & \mathbf{\thinspace\thinspace\thinspace\thinspace0}\\
-\mathbf{I} & \thinspace\thinspace\mathbf{\thinspace\thinspace0} & \thinspace\thinspace\mathbf{\thinspace\thinspace I}
\end{array}\right].
\end{equation}

We know from \cite{OmuraKailath1965,ParreiraBermudezRichardEtAl2012}
that the moment generating function of the quadratic form
$z=\boldsymbol{y}^{\text{T}}\mathbf{Q}\boldsymbol{y}$, where
$\boldsymbol{y}$ is a zero-mean Gaussian vector with covariance
matrix $\mathbf{R}_{y}$ is given by

\begin{equation}
\mathbb{E}\left[e^{sz}\right]=\mbox{det}\left\{ \mathbf{I}-2s\mathbf{Q}\mathbf{R}_{y}\right\} ^{-\frac{1}{2}}.
\end{equation}

The last equation allows us to compute the entries of the correlation matrix $\mathbf{R}_{kk}$ for the Gaussian kernel. Each element is given by

\begin{equation}
\left[\mathbf{R}_{kk}\right]_{jl}=\begin{cases}
r_{md}=\mbox{det}\left\{ \mathbf{I}_{2}-2\mathbf{Q}_{2}\mathbf{R}_{2}/\nu^{2}\right\} ^{-\frac{1}{2}} & j=l\\
r_{od}=\mbox{det}\left\{ \mathbf{I}_{3}-\mathbf{Q}_{3}\mathbf{R}_{3}/\nu^{2}\right\} ^{-\frac{1}{2}} & j\neq l
\end{cases}.
\end{equation}


Let us define the coefficients-error vector defined by
\begin{equation}
\boldsymbol{v}\left(i\right)=\boldsymbol{a}\left(i\right)-\boldsymbol{a}_{o}.\label{eq:kernel coeferror vector}
\end{equation}

The second-order moments of the coefficients are related to the MSE
through \cite{Sayed2008}
\begin{equation}
J_{\rm ms}\left(i\right)=J_{\rm min}+tr\left\{
\mathbf{R}_{kk}\mathbf{C}_{\boldsymbol{v}}\left(i\right)\right\} ,
\label{eq:mse}
\end{equation}
where
$\mathbf{C}_{\boldsymbol{v}}\left(i\right)=\mathbb{E}\left[\boldsymbol{v}\left(i\right)\boldsymbol{v}^{\text{T}}\left(i\right)\right]$.
This means that for studying the MSE behavior we need a model for
$\mathbf{C}_{\boldsymbol{v}}\left(i\right)$. In this section, we
derive an analytical model that describes the behavior of
$\mathbf{C}_{\boldsymbol{v}}\left(i\right)$ for the proposed
NLR-SM-KNLMS algorithm.

The update equation for the coefficients of the system is given by
\begin{equation}
\boldsymbol{a}\left(i+1\right)=\boldsymbol{a}\left(i\right)+\mu\left(i\right)e\left(i\right)\boldsymbol{\kappa_{\delta}}\left(i\right),\label{eq:update SM-KNLMS}
\end{equation}
where
\begin{equation}
\mu\left(i\right)=\begin{cases}
1-\frac{\gamma}{\left|e\left(i\right)\right|} & \left|e\left(i\right)\right|>\gamma,\\
0 & {\rm otherwise}.
\end{cases}
\end{equation}

Subtracting $\boldsymbol{a_{o}}$ from equation \eqref{eq:update SM-KNLMS}, we obtain the weight error vector update equation:
\begin{equation}
\boldsymbol{v}\left(i+1\right)=\boldsymbol{v}\left(i\right)+\mu\left(i\right)e\left(i\right)\boldsymbol{\kappa_{\delta}}\left(i\right).
\end{equation}

The estimation error may now be rewritten as follows:
\begin{align}
e\left(i\right) = & d\left(i\right)-\boldsymbol{\kappa}_{\boldsymbol{\delta}}^{\text{T}}\left(i\right)\boldsymbol{a}\left(i\right)\nonumber\\
 = & d\left(i\right)-\boldsymbol{\kappa}_{\boldsymbol{\delta}}^{\text{T}}
 \left(i\right)\boldsymbol{v}\left(i\right)-\boldsymbol{\kappa}_{\boldsymbol{\delta}}^{\text{T}}\left(i\right)\boldsymbol{a}_{o}.
\end{align}

The optimum error is given by
\begin{equation}
e_{o}\left(i\right)=d\left(i\right)-\boldsymbol{\kappa}_{\boldsymbol{\delta}}^{\text{T}}\left(i\right)\boldsymbol{a}_{o}.\label{eq:optimum error}
\end{equation}

It follows that
\begin{equation}
e\left(i\right)=e_{o}\left(i\right)-\boldsymbol{\kappa}_{\boldsymbol{\delta}}^{\text{T}}\left(i\right)\boldsymbol{v}\left(i\right).\label{eq:error in terms of optimum error}
\end{equation}

We may represent equation \eqref{eq:update SM-KNLMS} by
\begin{equation}
\boldsymbol{a}\left(i+1\right)=\boldsymbol{a}\left(i\right)+P_{\rm
up}\left(1-\frac{\gamma}{\left|e\left(i\right)\right|}\right)e\left(i\right)\boldsymbol{\kappa_{\delta}}\left(i\right),
\end{equation}
where $P_{\rm up} = {\rm Pr}(|e(i)|> \gamma)=2
Q\left(\frac{\gamma}{\sigma_e}\right)$ denotes the probability of
update of the set-membership algorithm \cite{LamareDiniz2013} and
$\sigma_e$ is the standard deviation of a Gaussian random variable
associated with the error.

Subtracting $\boldsymbol{a_{o}}$ from the last equation yields
\begin{align}
\boldsymbol{v}\left(i+1\right)  = & \boldsymbol{v}\left(i\right)+P_{\rm up}\left(1-\frac{\gamma}{\left|e\left(i\right)\right|}\right)e\left(i\right)
\boldsymbol{\kappa_{\delta}}\left(i\right)\nonumber\\
  = & \boldsymbol{v}\left(i\right)+P_{\rm up} e\left(i\right)\boldsymbol{\kappa_{\delta}}\left(i\right)\nonumber\\&-\gamma
  P_{\rm up}\mbox{sgn}\left(e\left(i\right)\right)\boldsymbol{\kappa_{\delta}}\left(i\right),
\end{align}

Replacing \eqref{eq:error in terms of optimum error} in the equation
above we obtain
\begin{align}
\boldsymbol{v}\left(i+1\right)  = & \boldsymbol{v}\left(i\right)+P_{\rm up}\left(e_{o}\left(i\right)-\boldsymbol{\kappa}_{\boldsymbol{\delta}}^{\text{T}}\left(i\right)\boldsymbol{v}\left(i\right)\right)\boldsymbol{\kappa_{\delta}}\left(i\right)\nonumber \\
  & -\gamma P_{\rm up}\mbox{sgn}\left(e\left(i\right)\right)\boldsymbol{\kappa_{\delta}}\left(i\right)\nonumber\\
  = & \boldsymbol{v}\left(i\right)+P_{\rm up}e_{o}\left(i\right)\boldsymbol{\kappa_{\delta}}\left(i\right)\nonumber\\&-P_{\rm up}\boldsymbol{\kappa_{\delta}}\left(i\right)\boldsymbol{\kappa}_{\boldsymbol{\delta}}^{\text{T}}\left(i\right)\boldsymbol{v}\left(i\right)\nonumber \\
   & -\gamma P_{\rm up}\mbox{sgn}\left(e\left(i\right)\right)\boldsymbol{\kappa_{\delta}}\left(i\right).\label{eq:v evolution}
\end{align}

Post-multiplying equation \eqref{eq:v evolution} by its transpose and taking the expected value leads to:

\begin{align}
\mathbf{C}_{\boldsymbol{v}}\left(i+1\right)= & \mathbf{C}_{\boldsymbol{v}}\left(i\right)+P_{\rm up}\mathbb{E}\left[e_{o}\left(i\right)\boldsymbol{v}\left(i\right)\boldsymbol{\kappa}_{\boldsymbol{\delta}}^{\text{T}}\left(i\right)\right]\nonumber\\
&-P_{\rm up}\mathbb{E}\left[\boldsymbol{v}\left(i\right)\boldsymbol{v}^{\text{T}}\left(i\right)\boldsymbol{\kappa_{\delta}}\left(i\right)\boldsymbol{\kappa}_{\boldsymbol{\delta}}^{\text{T}}\left(i\right)\right]\nonumber \\
 & -P_{\rm up}\gamma\mathbb{E}\left[\mbox{sgn}\left(e\left(i\right)\right)\boldsymbol{v}\left(i\right)\boldsymbol{\kappa}_{\boldsymbol{\delta}}^{\text{T}}\left(i\right)\right]\nonumber\\&+P_{\rm up}\mathbb{E}\left[e_{o}\left(i\right)\boldsymbol{\kappa_{\delta}}\left(i\right)\boldsymbol{v}^{\text{T}}\left(i\right)\right]\nonumber \\
  & +P_{\rm up}^2\mathbb{E}\left[e_{o}^{2}\left(i\right)\boldsymbol{\kappa_{\delta}}\left(i\right)\boldsymbol{\kappa}_{\boldsymbol{\delta}}^{\text{T}}\left(i\right)\right]\nonumber\\
  &-P_{\rm up}^2\mathbb{E}\left[e_{o}\left(i\right)\boldsymbol{\kappa_{\delta}}\left(i\right)\boldsymbol{v}^{\text{T}}\left(i\right)\boldsymbol{\kappa_{\delta}}\left(i\right)\boldsymbol{\kappa}_{\boldsymbol{\delta}}^{\text{T}}\left(i\right)\right]\nonumber \\
 & -2P_{\rm up}^2\gamma\mathbb{E}\left[e_{o}\left(i\right)\mbox{sgn}\left(e\left(i\right)\right)\boldsymbol{\kappa_{\delta}}\left(i\right)\boldsymbol{\kappa}_{\boldsymbol{\delta}}^{\text{T}}\left(i\right)\right]\nonumber \\
 & -P_{\rm up}\mathbb{E}\left[\boldsymbol{\kappa_{\delta}}\left(i\right)\boldsymbol{\kappa}_{\boldsymbol{\delta}}^{\text{T}}\left(i\right)\boldsymbol{v}\left(i\right)\boldsymbol{v}^{\text{T}}\left(i\right)\right]\nonumber \\
 & -P_{\rm up}^2\mathbb{E}\left[e_{o}\left(i\right)\boldsymbol{\kappa_{\delta}}\left(i\right)\boldsymbol{\kappa}_{\boldsymbol{\delta}}^{\text{T}}\left(i\right)\boldsymbol{v}\left(i\right)\boldsymbol{\kappa}_{\boldsymbol{\delta}}^{\text{T}}\left(i\right)\right]\nonumber \\
 & +P_{\rm up}^2\mathbb{E}\left[\boldsymbol{\kappa_{\delta}}\left(i\right)\boldsymbol{\kappa}_{\boldsymbol{\delta}}^{\text{T}}\left(i\right)\boldsymbol{v}\left(i\right)\boldsymbol{v}^{\text{T}}\left(i\right)\boldsymbol{\kappa_{\delta}}\left(i\right)\boldsymbol{\kappa}_{\boldsymbol{\delta}}^{\text{T}}\left(i\right)\right]\nonumber \\
 & +P_{\rm up}^2\gamma\mathbb{E}\left[\mbox{sgn}\left(e\left(i\right)\right)\boldsymbol{\kappa_{\delta}}\left(i\right)\boldsymbol{\kappa}_{\boldsymbol{\delta}}^{\text{T}}\left(i\right)\boldsymbol{v}\left(i\right)\boldsymbol{\kappa}_{\boldsymbol{\delta}}^{\text{T}}\left(i\right)\right]\nonumber \\
 & -P_{\rm up}\gamma\mathbb{E}\left[\mbox{sgn}\left(e\left(i\right)\right)\boldsymbol{\kappa_{\delta}}\left(i\right)\boldsymbol{v}^{\text{T}}\left(i\right)\right]\nonumber \\
 & +P_{\rm up}^2\gamma\mathbb{E}\left[\mbox{sgn}\left(e\left(i\right)\right)\boldsymbol{\kappa_{\delta}}\left(i\right)\boldsymbol{v}^{\text{T}}\left(i\right)\boldsymbol{\kappa_{\delta}}\left(i\right)\boldsymbol{\kappa}_{\boldsymbol{\delta}}^{\text{T}}\left(i\right)\right]\nonumber \\
 & +P_{\rm up}^2\gamma^{2}\mathbb{E}\left[\mbox{sgn}^{2}\left(e\left(i\right)\right)\boldsymbol{\kappa_{\delta}}\left(i\right)\boldsymbol{\kappa}_{\boldsymbol{\delta}}^{\text{T}}\left(i\right)\right].\label{eq:cv evolution uno}
\end{align}

Let us define $\mathbf{T}\left(i\right)=\mathbb{E}\left[\boldsymbol{\kappa_{\delta}}\left(i\right)\boldsymbol{\kappa}_{\boldsymbol{\delta}}^{\text{T}}\left(i\right)\boldsymbol{v}\left(i\right)\boldsymbol{v}^{\text{T}}\left(i\right)\boldsymbol{\kappa_{\delta}}\left(i\right)\boldsymbol{\kappa}_{\boldsymbol{\delta}}^{\text{T}}\left(i\right)\right]$ to simplify the notation. Assuming that the inputs and the coefficients are statistically independent, then the following expected values are reduced to
\begin{align}
\mathbb{E}\left[\boldsymbol{\kappa_{\delta}}\left(i\right)\boldsymbol{\kappa}_{\boldsymbol{\delta}}^{\text{T}}\left(i\right)\boldsymbol{v}\left(i\right)\boldsymbol{v}^{\text{T}}\left(i\right)\right]  = & \mathbf{R}_{kk}\mathbf{C}_{\boldsymbol{v}}\left(i\right),\label{eq:expectedval1}\\
\mathbb{E}\left[\boldsymbol{v}\left(i\right)\boldsymbol{v}^{\text{T}}\left(i\right)\boldsymbol{\kappa_{\delta}}\left(i\right)\boldsymbol{\kappa}_{\boldsymbol{\delta}}^{\text{T}}\left(i\right)\right]  = & \mathbf{C}_{\boldsymbol{v}}\left(i\right)\mathbf{R}_{kk}\label{eq:expectedval2}.
\end{align}

Let us also suppose that the optimum error is independent from the
kernelized inputs. This assumption leads us to:
\begin{align}
\mathbb{E}\left[e_{o}^{2}\left(i\right)\boldsymbol{\kappa_{\delta}}\left(i\right)\boldsymbol{\kappa}_{\boldsymbol{\delta}}^{\text{T}}\left(i\right)\right]  \approx & \mathbb{E}\left[e_{o}^{2}\left(i\right)\right]\mathbb{E}\left[\boldsymbol{\kappa_{\delta}}\left(i\right)\boldsymbol{\kappa}_{\boldsymbol{\delta}}^{\text{T}}\left(i\right)\right]\nonumber\\
  \approx & J_{\rm min}\mathbf{R}_{kk}\label{eq:expectedval3},
\end{align}
and
\begin{align}
\mathbb{E}\left[\mbox{sgn}^{2}\left(e\left(i\right)\right)\boldsymbol{\kappa_{\delta}}\left(i\right)\boldsymbol{\kappa}_{\boldsymbol{\delta}}^{\text{T}}\left(i\right)\right]  \approx & \mathbb{E}\left[\mbox{sgn}^{2}\left(e\left(i\right)\right)\right]\mathbb{E}\left[\boldsymbol{\kappa_{\delta}}\left(i\right)\boldsymbol{\kappa}_{\boldsymbol{\delta}}^{\text{T}}\left(i\right)\right]\nonumber\\
  \approx & \mathbf{R}_{kk}\label{eq:expectedval4}.
\end{align}

By the orthogonality principle, we obtain:
\begin{align}
\mathbb{E}\left[e_{o}\left(i\right)\boldsymbol{\kappa_{\delta}}\left(i\right)\boldsymbol{v}^{\text{T}}\left(i\right)\right] \approx & \mathbb{E}\left[e_{o}\left(i\right)\boldsymbol{\kappa_{\delta}}\left(i\right)\right]\mathbb{E}\left[\boldsymbol{v}^{\text{T}}\left(i\right)\right]\nonumber\\
  \approx & \mathbf{0}\label{eq:expectedval5},
\end{align}

Let us also apply the orthogonality principle in the following expected value:
\begin{align}
\mathbb{E}&\left[e_{o}\left(i\right)\boldsymbol{\kappa_{\delta}}\left(i\right)\boldsymbol{v}^{\text{T}}\left(i\right)\boldsymbol{\kappa_{\delta}}\left(i\right)\boldsymbol{\kappa}_{\boldsymbol{\delta}}^{\text{T}}\left(i\right)\right]\nonumber\\& = \mathbb{E}\left[\boldsymbol{v}^{\text{T}}\left(i\right)e_{o}\left(i\right)\boldsymbol{\kappa_{\delta}}\left(i\right)\boldsymbol{\kappa_{\delta}}\left(i\right)\boldsymbol{\kappa}_{\boldsymbol{\delta}}^{\text{T}}\left(i\right)\right]\nonumber\\
 &\approx\mathbb{E}\left[\boldsymbol{v}^{\text{T}}\left(i\right)\right]\mathbb{E}\left[e_{o}\left(i\right)\boldsymbol{\kappa_{\delta}}\left(i\right)\right]\mathbb{E}\left[\boldsymbol{\kappa_{\delta}}\left(i\right)\boldsymbol{\kappa}_{\boldsymbol{\delta}}^{\text{T}}\left(i\right)\right]\nonumber\\
 &\approx\mathbf{0}\label{eq:expectedval6}.
\end{align}

With the results of equations \eqref{eq:expectedval1},\eqref{eq:expectedval2}, \eqref{eq:expectedval3}, \eqref{eq:expectedval4}, \eqref{eq:expectedval5} and \eqref{eq:expectedval6}, equation \eqref{eq:cv evolution uno} is reduced to:
\begin{align}
\mathbf{C}_{\boldsymbol{v}}\left(i+1\right) = & \mathbf{C}_{\boldsymbol{v}}\left(i\right)-P_{\rm up}\mathbf{C}_{\boldsymbol{v}}\left(i\right)\mathbf{R}_{kk}\nonumber\\&-P_{\rm up}\gamma\mathbb{E}\left[\mbox{sgn}\left(e\left(i\right)\right)\boldsymbol{v}\left(i\right)\boldsymbol{\kappa}_{\boldsymbol{\delta}}^{\text{T}}\left(i\right)\right]\nonumber \\
  & +P_{\rm up}^2J_{\rm min}\mathbf{R}_{kk}\nonumber\\&-2P_{\rm up}^2\gamma\mathbb{E}\left[e_{o}\left(i\right)\mbox{sgn}\left(e\left(i\right)\right)\boldsymbol{\kappa_{\delta}}\left(i\right)\boldsymbol{\kappa}_{\boldsymbol{\delta}}^{\text{T}}\left(i\right)\right]\nonumber\\
  & -P_{\rm up}\mathbf{R}_{kk}\mathbf{C}_{\boldsymbol{v}}\left(i\right)+P_{\rm up}^2\mathbf{T}\left(i\right)\nonumber\\
  & +P_{\rm up}^2\gamma\mathbb{E}\left[\mbox{sgn}\left(e\left(i\right)\right)\boldsymbol{\kappa_{\delta}}\left(i\right)\boldsymbol{\kappa}_{\boldsymbol{\delta}}^{\text{T}}\left(i\right)\boldsymbol{v}\left(i\right)\boldsymbol{\kappa}_{\boldsymbol{\delta}}^{\text{T}}\left(i\right)\right]\nonumber\\
  & -P_{\rm up}\gamma\mathbb{E}\left[\mbox{sgn}\left(e\left(i\right)\right)\boldsymbol{\kappa_{\delta}}\left(i\right)\boldsymbol{v}^{\text{T}}\left(i\right)\right]\nonumber\\
  & +P_{\rm up}^2\gamma\mathbb{E}\left[\mbox{sgn}\left(e\left(i\right)\right)\boldsymbol{\kappa_{\delta}}\left(i\right)\boldsymbol{v}^{\text{T}}\left(i\right)\boldsymbol{\kappa_{\delta}}\left(i\right)\boldsymbol{\kappa}_{\boldsymbol{\delta}}^{\text{T}}\left(i\right)\right]\nonumber\\
  & +P_{\rm up}^2\gamma^{2}\mathbf{R}_{kk},\label{eq:cv evolution dosa}
\end{align}
The remaining expected values of \eqref{eq:cv evolution dosa} can be
computed using  {Price's theorem \cite{Price}}. For the ninth term,
the expected value may be approximated as follows:

\begin{align}
\mathbb{E}&\left[\mbox{sgn}\left(e\left(i\right)\right)\boldsymbol{\kappa_{\delta}}\left(i\right)\boldsymbol{v}^{\text{T}}\left(i\right)\right] \nonumber\\&\approx \sqrt{\frac{2}{\pi\sigma_{e}^{2}}}\mathbb{E}\left[e\left(i\right)\boldsymbol{\kappa_{\delta}}\left(i\right)\boldsymbol{v}^{\text{T}}\left(i\right)\right]\nonumber\\
 & \approx \sqrt{\frac{2}{\pi\sigma_{e}^{2}}}\mathbb{E}\left[\left(e_{o}\left(i\right)-\boldsymbol{\kappa}_{\boldsymbol{\delta}}^{\text{T}}\left(i\right)\boldsymbol{v}\left(i\right)\right)\boldsymbol{\kappa_{\delta}}\left(i\right)\boldsymbol{v}^{\text{T}}\left(i\right)\right]\nonumber\\
 & \approx -\sqrt{\frac{2}{\pi\sigma_{e}^{2}}}\mathbb{E}\left[\boldsymbol{\kappa_{\delta}}\left(i\right)\boldsymbol{\kappa}_{\boldsymbol{\delta}}^{\text{T}}\left(i\right)\boldsymbol{v}\left(i\right)\boldsymbol{v}^{\text{T}}\left(i\right)\right]\nonumber\\
 & \approx -\sqrt{\frac{2}{\pi\sigma_{e}^{2}}}\mathbf{R}_{kk}\mathbf{C}_{\boldsymbol{v}}\left(i\right)\label{eq:exvalprice1}.
\end{align}

Calculating the third term of equation \eqref{eq:cv evolution dosa}, we obtain

\begin{align}
\mathbb{E}\left[\mbox{sgn}\left(e\left(i\right)\right)\boldsymbol{v}\left(i\right)\boldsymbol{\kappa}_{\boldsymbol{\delta}}^{\text{T}}\left(i\right)\right]  \approx & \sqrt{\frac{2}{\pi\sigma_{e}^{2}}}\mathbb{E}\left[e\left(i\right)\boldsymbol{v}\left(i\right)\boldsymbol{\kappa}_{\boldsymbol{\delta}}^{\text{T}}\left(i\right)\right]\nonumber\\
  \approx & -\sqrt{\frac{2}{\pi\sigma_{e}^{2}}}\mathbf{C}_{\boldsymbol{v}}\left(i\right)\mathbf{R}_{kk}.\label{eq:exvalprice2}
\end{align}

The sixth term of equation \eqref{eq:cv evolution dosa} is given by
\begin{align}
\mathbb{E}&\left[e_{o}\left(i\right)\mbox{sgn}\left(e\left(i\right)\right)\boldsymbol{\kappa_{\delta}}\left(i\right)\boldsymbol{\kappa}_{\boldsymbol{\delta}}^{\text{T}}\left(i\right)\right] \nonumber\\&\approx \sqrt{\frac{2}{\pi\sigma_{e}^{2}}}\mathbb{E}\left[e_{o}\left(i\right)e\left(i\right)\boldsymbol{\kappa_{\delta}}\left(i\right)\boldsymbol{\kappa}_{\boldsymbol{\delta}}^{\text{T}}\left(i\right)\right]\nonumber\\
 &\approx \sqrt{\frac{2}{\pi\sigma_{e}^{2}}}\mathbb{E}\left[e_{o}^{2}\left(i\right)\boldsymbol{\kappa_{\delta}}\left(i\right)\boldsymbol{\kappa}_{\boldsymbol{\delta}}^{\text{T}}\left(i\right)\right]\nonumber \\
 &~~~~-\sqrt{\frac{2}{\pi\sigma_{e}^{2}}}\mathbb{E}\left[e_{o}\left(i\right)\boldsymbol{\kappa}_{\boldsymbol{\delta}}^{\text{T}}\left(i\right)\boldsymbol{v}\left(i\right)\boldsymbol{\kappa_{\delta}}\left(i\right)\boldsymbol{\kappa}_{\boldsymbol{\delta}}^{\text{T}}\left(i\right)\right]\nonumber\\
& \approx  \sqrt{\frac{2}{\pi\sigma_{e}^{2}}}J_{\rm
min}\mathbf{R}_{kk}.\label{eq:exvalprice3}
\end{align}

Finally, the eighth and the tenth terms can be computed by
\begin{align}
\mathbb{E}&\left[\mbox{sgn}\left(e\left(i\right)\right)\boldsymbol{\kappa_{\delta}}\left(i\right)\boldsymbol{\kappa}_{\boldsymbol{\delta}}^{\text{T}}\left(i\right)\boldsymbol{v}\left(i\right)\boldsymbol{\kappa}_{\boldsymbol{\delta}}^{\text{T}}\left(i\right)\right]\nonumber\\ &\approx\sqrt{\frac{2}{\pi\sigma_{e}^{2}}}\mathbb{E}\left[e\left(i\right)\boldsymbol{\kappa_{\delta}}\left(i\right)\boldsymbol{\kappa}_{\boldsymbol{\delta}}^{\text{T}}\left(i\right)\boldsymbol{v}\left(i\right)\boldsymbol{\kappa}_{\boldsymbol{\delta}}^{\text{T}}\left(i\right)\right]\nonumber\\
&\approx-\sqrt{\frac{2}{\pi\sigma_{e}^{2}}}\mathbf{T}\left(i\right).\label{eq:exvalprice4}
\end{align}

 {The results obtained in \eqref{eq:exvalprice1},
\eqref{eq:exvalprice2}, \eqref{eq:exvalprice3} and
\eqref{eq:exvalprice4} shall turn \eqref{eq:cv evolution dosa}
into:}
\begin{align}
\mathbf{C}_{\boldsymbol{v}}\left(i+1\right)  = & \mathbf{C}_{\boldsymbol{v}}\left(i\right)-P_{\rm up}\mathbf{C}_{\boldsymbol{v}}\left(i\right)\mathbf{R}_{kk}\nonumber\\&+P_{\rm up}\gamma\sqrt{\frac{2}{\pi\sigma_{e}^{2}}}\mathbf{C}_{\boldsymbol{v}}\left(i\right)\mathbf{R}_{kk}\nonumber \\
   & +P_{\rm up}^2J_{\rm min}\mathbf{R}_{kk}-2P_{\rm up}^2\gamma\sqrt{\frac{2}{\pi\sigma_{e}^{2}}}J_{\rm min}\mathbf{R}_{kk}\nonumber \\
   & -P_{\rm up}\mathbf{R}_{kk}\mathbf{C}_{\boldsymbol{v}}\left(i\right)+P_{\rm up}^2\mathbf{T}\left(i\right)\nonumber \\
   & -2P_{\rm up}^2\gamma\sqrt{\frac{2}{\pi\sigma_{e}^{2}}}\mathbf{T}\left(i\right)\nonumber\\&+P_{\rm up}\gamma\sqrt{\frac{2}{\pi\sigma_{e}^{2}}}\mathbf{R}_{kk}\mathbf{C}_{\boldsymbol{v}}\left(i\right)\nonumber \\
   & +P_{\rm up}^2\gamma^{2}\mathbf{R}_{kk}.
\end{align}

Factorizing the common terms of the last equation, we get the following recursion for $\mathbf{C}_{\boldsymbol{v}}\left(i+1\right)$:
\begin{align}
\mathbf{C}_{\boldsymbol{v}}\left(i+1\right)= & \mathbf{C}_{\boldsymbol{v}}\left(i\right)+P_{\rm up}^2\left(1-2\gamma\sqrt{\frac{2}{\pi\sigma_{e}^{2}}}\right)\left(J_{\rm min}\mathbf{R}_{kk}+\mathbf{T}\left(i\right)\right)\nonumber \\
 & +P_{\rm up}\left(\gamma\sqrt{\frac{2}{\pi\sigma_{e}^{2}}}-1\right)\left(\mathbf{C}_{\boldsymbol{v}}\left(i\right)\mathbf{R}_{kk}+\mathbf{R}_{kk}\mathbf{C}_{\boldsymbol{v}}\right)\nonumber\\&+P_{\rm up}^2\gamma^{2}\mathbf{R}_{kk}.\label{eq:cv evolution dos}
\end{align}

The authors of \cite{ParreiraBermudezRichardEtAl2012} proved that
the elements of $\mathbf{T}\left(i\right)$ are given by
\begin{align}
\left[\mathbf{T}\left(i\right)\right]_{jj}  = & \mathop{\sum_{l=1}^{M}}_{l\neq j} \left\{ 2\mu_{2}\left[\mathbf{C}_{\boldsymbol{v}}\left(i\right)\right]_{jl}+\mu_{3}\left[\mathbf{C}_{\boldsymbol{v}}\left(i\right)\right]_{ll}+\mu_{4}\mathop{\sum_{p=1}^{M}}_{p\neq \left\{ j,l\right\}}\left[\mathbf{C}_{\boldsymbol{v}}\left(i\right)\right]_{lp}\right\} \nonumber \\
  & +\mu_{1}\left[\mathbf{C}_{\boldsymbol{v}}\left(i\right)\right]_{jj},\label{eq:T recursivo digonal}
\end{align}

for the main diagonal elements and  
\begin{align}
\left[\mathbf{T}\left(i\right)\right]_{jk} = & \mu_{2}\left(\left[\mathbf{C}_{\boldsymbol{v}}\left(i\right)\right]_{jj}+\left[\mathbf{C}_{\boldsymbol{v}}\left(i\right)\right]_{kk}\right)+2\mu_{3}\left[\mathbf{C}_{\boldsymbol{v}}\left(i\right)\right]_{jk}\nonumber \\
  & +\mathop{\sum_{l=1}^{M}}_{l\neq\left\{ j,k\right\} }\Biggl\{2\mu_{4}\left[\mathbf{C}_{\boldsymbol{v}}\left(i\right)\right]_{kl}+2\mu_{4}\left[\mathbf{C}_{\boldsymbol{v}}\left(i\right)\right]_{jl}+\mu_{4}\left[\mathbf{C}_{\boldsymbol{v}}\left(i\right)\right]_{ll}\nonumber \\
  & \left.+\mu_{5}\mathop{\sum_{p=1}^{M}}_{p\neq\left\{ j,k,l\right\} }\left[\mathbf{C}_{\boldsymbol{v}}\left(i\right)\right]_{lp}\right\},\label{eq:T recursivo off diagonal}
\end{align}
for  the off-diagonal entries, where $\mu_i$ is defined by
\begin{align}
\mu_{1}=&\mbox{det}\left\{ \mathbf{I}_{2}-4\mathbf{Q}_{2}\mathbf{R}_{2}/\nu^{2}\right\} ^{-\frac{1}{2}},\\
\mu_{2}=&\mbox{det}\left\{ \mathbf{I}_{3}-\mathbf{Q}_{3'}\mathbf{R}_{3}/\nu^{2}\right\} ^{-\frac{1}{2}},\\
\mu_{3}=&\mbox{det}\left\{ \mathbf{I}_{3}-2\mathbf{Q}_{3'}\mathbf{R}_{3}/\nu^{2}\right\} ^{-\frac{1}{2}},\\
\mu_{4}=&\mbox{det}\left\{ \mathbf{I}_{4}-2\mathbf{Q}_{4}\mathbf{R}_{4}/\nu^{2}\right\} ^{-\frac{1}{2}},\\
\mu_{5}=&\mbox{det}\left\{ \mathbf{I}_{5}-2\mathbf{Q}_{5}\mathbf{R}_{5}/\nu^{2}\right\} ^{-\frac{1}{2}},
\end{align}

and the matrices $\mathbf{Q}_i$ are defined by
\begin{equation}
\mathbf{Q}_{3'}=\left[\begin{array}{ccc}
\thinspace\thinspace\thinspace\thinspace4\mathbf{I} & -3\mathbf{I} & -\mathbf{I}\\
-3\mathbf{I} & \thinspace\thinspace\thinspace\thinspace3\mathbf{I} & \mathbf{\thinspace\thinspace\thinspace\thinspace0}\\
-\mathbf{I} & \thinspace\thinspace\mathbf{\thinspace\thinspace\thinspace\thinspace0} & \thinspace\thinspace\mathbf{\thinspace\thinspace I}
\end{array}\right],
\end{equation}
\begin{equation}
\mathbf{Q}_{4}=\left[\begin{array}{cccc}
\thinspace\thinspace\thinspace\thinspace4\mathbf{I} & -2\mathbf{I} & -\mathbf{I} & -\mathbf{I}\\
-2\mathbf{I} & \thinspace\thinspace\thinspace\thinspace2\mathbf{I} & \mathbf{\thinspace\thinspace\thinspace\thinspace0} & \mathbf{\thinspace\thinspace\thinspace\thinspace0}\\
-\mathbf{I} & \thinspace\thinspace\mathbf{\thinspace\thinspace\thinspace\thinspace0} & \thinspace\thinspace\mathbf{\thinspace\thinspace I} & \thinspace\thinspace\mathbf{\thinspace\thinspace0}\\
-\mathbf{I} & \thinspace\thinspace\mathbf{\thinspace\thinspace\thinspace\thinspace0} & \thinspace\thinspace\thinspace\thinspace\mathbf{0} & \thinspace\thinspace\thinspace\thinspace\mathbf{I}
\end{array}\right],
\end{equation}
\\
\begin{equation}
\mathbf{Q}_{5}=\left[\begin{array}{ccccc}
\thinspace\thinspace\thinspace\thinspace4\mathbf{I} & -\mathbf{I} & -\mathbf{I} & -\mathbf{I} & -\mathbf{I}\\
-\mathbf{I} & \thinspace\thinspace\thinspace\thinspace\thinspace\mathbf{I} & \mathbf{\thinspace\thinspace\thinspace\thinspace0} & \mathbf{\thinspace\thinspace\thinspace\thinspace0} & \mathbf{\thinspace\thinspace\thinspace\thinspace0}\\
-\mathbf{I} & \thinspace\thinspace\mathbf{\thinspace\thinspace\thinspace0} & \thinspace\thinspace\mathbf{\thinspace\thinspace I} & \thinspace\thinspace\mathbf{\thinspace\thinspace0} & \mathbf{\thinspace\thinspace\thinspace\thinspace0}\\
-\mathbf{I} & \thinspace\thinspace\mathbf{\thinspace\thinspace\thinspace0} & \thinspace\thinspace\thinspace\thinspace\mathbf{0} & \thinspace\thinspace\thinspace\thinspace\mathbf{I} & \mathbf{\thinspace\thinspace\thinspace\thinspace0}\\
-\mathbf{I} & \mathbf{\thinspace\thinspace\thinspace\thinspace\thinspace0} & \mathbf{\thinspace\thinspace\thinspace\thinspace0} & \mathbf{\thinspace\thinspace\thinspace\thinspace0} & \thinspace\thinspace\thinspace\thinspace\mathbf{I}
\end{array}\right].
\end{equation}

Replacing equations \eqref{eq:T recursivo digonal} and \eqref{eq:T recursivo off diagonal} into equation \eqref{eq:cv evolution dos} leads
us to a recursive expression for the entries of the autocorrelation
matrix $\mathbf{C}_{\boldsymbol{v}}\left(i\right)$:
\begin{align}
\left[\mathbf{C}_{\boldsymbol{v}}\left(i+1\right)\right]_{jj}= & \left(1+2P_{\rm up}ar_{md}+P_{\rm up}^2b\mu_{1}\right)\left[\mathbf{C}_{\boldsymbol{v}}\left(i\right)\right]_{jj}\nonumber \\
 & +P_{up}^{2}b\mu_{3}\mathop{\sum_{l=1}^{M}}_{l\neq j}\left[\mathbf{C}_{\boldsymbol{v}}\left(i\right)\right]_{ll}\nonumber \\
 & +\left(2P_{\rm up}^2\mu_{2}b+2P_{\rm up}ar_{od}\right)\mathop{\sum_{l=1}^{M}}_{l\neq j}\left[\mathbf{C}_{\boldsymbol{v}}\left(i\right)\right]_{jl}\nonumber \\
 & +P_{\rm up}^2\mu_{2}b\mu_{4}\mathop{\sum_{l=1}^{M}}_{l\neq j}\mathop{\sum_{p=1}^{M}}_{p\neq\left\{ j,l\right\} }\left[\mathbf{C}_{\boldsymbol{v}}\left(i\right)\right]_{lp}\nonumber \\
 & +\left(P_{\rm up}^2bJ_{\rm min}+P_{\rm up}^2\gamma^{2}\right)r_{md},
\end{align}

and for $j\neq k$
\begin{align}
\left[\mathbf{C}_{\boldsymbol{v}}\left(i+1\right)\right]_{jk} = & \left(1+2P_{\rm up}\alpha r_{md}
+2P_{\rm up}^2\beta\mu_{3}\right)\left[\mathbf{C}_{\boldsymbol{v}}\left(i\right)\right]_{jk}\nonumber \\
  & +P_{\rm up}^2\beta\mu_{4}\mathop{\sum_{l=1}^{M}}_{l\neq\left\{ j,k\right\} }\left[\mathbf{C}_{\boldsymbol{v}}\left(i\right)\right]_{ll}+\Biggl(P_{\rm up}^2\beta\mu_{2}\nonumber \\
  & +P_{\rm up}\alpha r_{od}\Biggr)\left(\left[\mathbf{C}_{\boldsymbol{v}}\left(i\right)\right]_{jj}+\left[\mathbf{C}_{\boldsymbol{v}}\left(i\right)\right]_{kk}\right)\nonumber \\
  & +\Biggl(2P_{\rm up}^2\beta\mu_{4}+P_{\rm up}\alpha r_{od}\Biggr)\mathop{\sum_{l=1}^{M}}_{l\neq\left\{ j,k\right\} }\Biggl(\left[\mathbf{C}_{\boldsymbol{v}}\left(i\right)\right]_{il}\nonumber \\
  & +\left[\mathbf{C}_{\boldsymbol{v}}\left(i\right)\right]_{jl}\Biggr)+P_{\rm up}^2\beta\mu_{5}\mathop{\sum_{l=1}^{M}}_{l\neq\left\{ j,k\right\} }\mathop{\sum_{p=1}^{M}}_{p\neq\left\{ j,k,l\right\} }\left[\mathbf{C}_{\boldsymbol{v}}\left(i\right)\right]_{lp}\nonumber \\
  & +\left(P_{\rm up}^2\beta J_{\rm min}+P_{\rm up}^2\gamma^{2}\right)r_{md},
\end{align}

where
\begin{align}
\alpha=&\gamma\sqrt{\frac{2}{\pi\sigma_{e}^{2}}}-1,\\
\beta=&1-2\gamma\sqrt{\frac{2}{\pi\sigma_{e}^{2}}}.
\end{align}
The entries of the autocorrelation matrix
$\mathbf{C}_{\boldsymbol{v}}\left(i\right)$ are then plugged in the
MSE expression in (\ref{eq:mse}).

\section{Simulations}

In this section we assess the performance of the SM-KNLMS algorithms
proposed. The Gaussian kernel was used in all the algorithms to
perform all the experiments.  {We have structured this section into
two parts: the first part deals with the identification of nonlinear
systems, whereas the second part examines time series prediction
problems.}

\subsection{System identification}

In the first example, we consider a system identification
application to compare the performance of the proposed SM-KNLMS
algorithms and to verify the theory developed in Section
\ref{sec:Statistical Analysis}. Let us consider the nonlinear
problem studied in
\cite{NarendraParthasarathy1990,Mandic2004,ParreiraBermudezRichardEtAl2012}
described by the recursion
\begin{equation}
d\left(i\right)=\frac{d\left(i-1\right)}{1+d\left(i-1\right)}+x^3\left(i-1\right).
\end{equation}

\begin{figure}[h]
\begin{centering}
\includegraphics[scale=0.5]{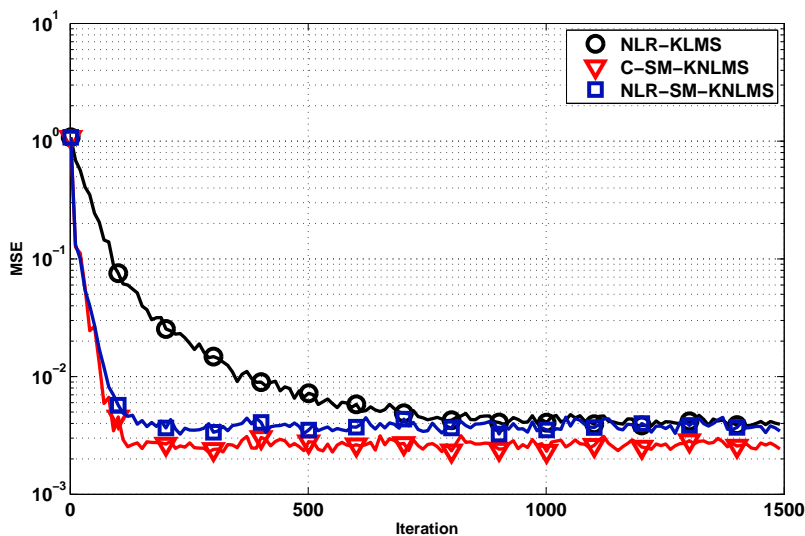}\caption{Performance comparison
of SM-KNLMS algorithms.} \label{fig:Performance-comparison-SM-KNLMS}
\par\end{centering}
\end{figure}

We compare the performance of the proposed algorithms with the KLMS
algorithm. The desired signal $d\left(i\right)$ was corrupted by
additive white Gaussian noise with zero-mean and variance
$\sigma^2_n=10^{-4}$ and the SNR was set to 20 dB. We have also
considered a fixed dictionary of length of $16$ in order to focus
solely on the performance of the gradient learning rules used by the
analyzed algorithms. At each iteration, the dictionary elements were
updated so that the oldest element added is replaced. To compute the
learning curve, a total of $500$ simulations were averaged, each one
with $1500$ iterations. The bandwidth of the Gaussian kernel was set
to 0.025. The threshold for the SM-KNLMS algorithms was set to
$\gamma=\sqrt{5}\sigma_n$. The result of this experiment is
presented in Fig. \ref{fig:Performance-comparison-SM-KNLMS}. The
results show that the C-SM-KNLMS algorithm slightly outperforms in
learning rate the NLR-SM-KNLMS algorithm and the nonlinear
regression-based KLMS (NLR-KLMS) algorithm. At steady state
C-SM-KNLMS and NLR-KNLMS tend to produce comparable results, which
means that the analytical formulas to predict the results of
NLR-KNLMS can be useful to have a prediction of the performance of
C-SM-NKLMS at steady state. For this reason we will consider the
C-SM-KNLMS algorithm for most examples except for those that show
analytical results and employ the NLR-SM-KNLMS algorithm, which is
the only one suitable for statistical analysis.

\begin{figure}[H]
\begin{centering}
\includegraphics[scale=0.5]{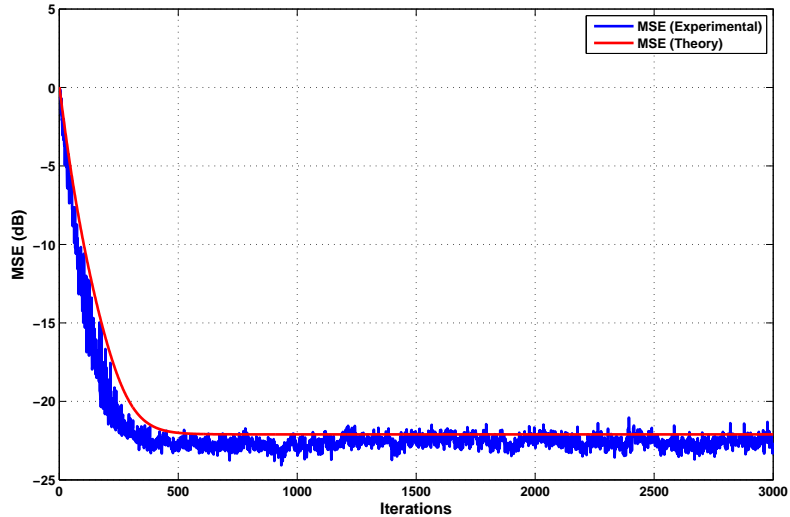}\caption{Transient behaviour of the SM-KNLMS algorithm.
Bandwidth=0.025} \label{fig:Performance-transient2}
\par\end{centering}
\end{figure}

In the second example we evaluate the transient behavior of the
NLR-SM-KNLMS. The input sequence $x\left(i\right)$ is independent
and identically Gaussian distributed with variance
$\sigma^2_x=0.15$. We have also considered a fixed dictionary of
length $16$. At each iteration, the dictionary elements were updated
so that the oldest element added is replaced. To compute the
learning curve, a total of $500$ simulations were averaged, each one
with $3000$ iterations. The bandwidth of the Gaussian kernel was set
to $0.025$ for Fig. \ref{fig:Performance-transient2} and the
threshold was set to $\gamma=\sqrt{10}\sigma_n$
to obtain the results in Fig.  \ref{fig:Performance-transient2}. 

{In the third example, we assess the performance of the NLR-SM-KNLMS
algorithm in a non-stationary environment. Particularly, we
investigate the case when a sudden change occurs the system model,
resulting in a different value of $\boldsymbol{\alpha_{o}}$. The two
systems studied are given by
\begin{equation}
d_{1}\left(i\right)=\frac{d\left(i-1\right)}{1+d\left(i-1\right)}+x^{3}\left(i-1\right),
\end{equation}
\begin{equation}
d_{2}\left(i\right)=x^{2}(i).
\end{equation}}
 {In particular, a total of $8000$ iterations were
made, where the first $4000$ iterations correspond to system
$d_{1}$. Then, the system becomes unstable for $50$ iterations where
$d(i)=d(i-1)+0.1$. The remaining iterations correspond to system
$d_{2}$. The output is corrupted by AWGN with standard deviation
equal to $\sigma_{n}=0.01$. The input follows a Gaussian
distribution with i.i.d samples and standard deviation given by
$\sigma_{x}=0.15$.}

\begin{figure}[H]
\begin{centering}
\includegraphics[scale=0.5]{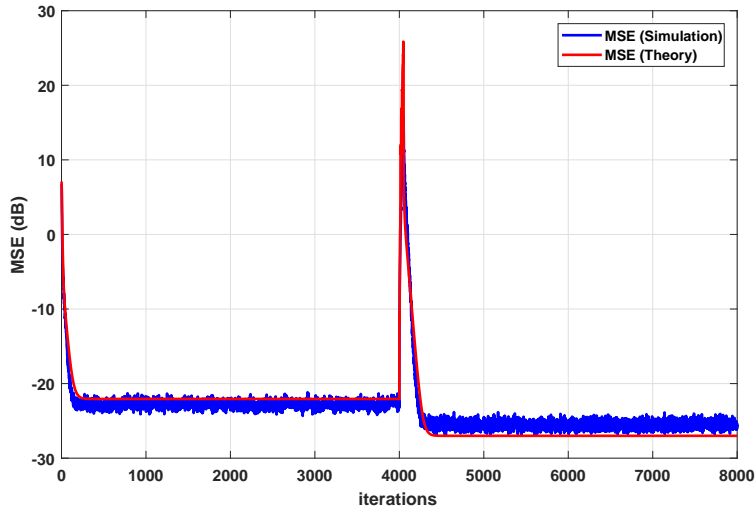}
\par\end{centering}
\caption{Performance of the NLR-SM-KNLMS algorithm in a
non-stationary environment.} \label{fig:change}
\end{figure}

 {From Fig. \ref{fig:change} we note that NLR-SM-KNLMS
is capable of tracking changes on the system and of converging to a
new solution in few iterations. The solution obtained for system
$d_{2}$ achieves a lower MSE because the correlation between the
mapped input and the desired signal is higher for this system. It is
also important to mention that the simulation result matches the
theoretical result.}

\begin{figure}[H]
\begin{centering}
\includegraphics[scale=0.5]{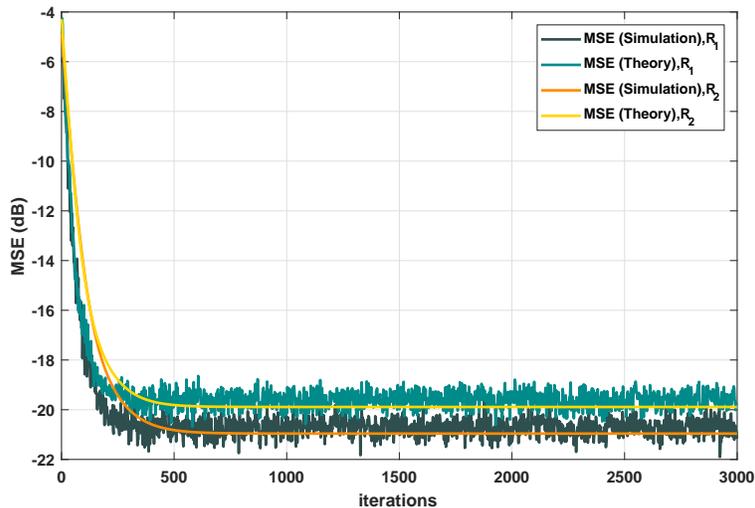}
\par\end{centering}
\caption{Performance of the NLR-SM-KNLMS algorithm with correlated
inputs.} \label{correlated}
\end{figure}

 {In the fourth experiment we assess the performance
of NLR-SM-KNLMS for correlated inputs. Let us consider two inputs,
$\mathbf{x}_{c_{1}}$ and $\mathbf{x}_{c_{2}}$ each one with three
different components i.e.
$\mathbf{x}_{c}\left(i\right)=\left[\begin{array}{ccc}
x_{c,1}\left(i\right) & x_{c,2}\left(i\right) &
x_{c,3}\left(i\right)\end{array}\right]^{T}$. The correlation of the
inputs satisfies}
\begin{equation}
x_{c_{1},2}\left(i\right)=0.5x_{c_{1},1}\left(i\right)+\delta_{x}\left(i\right)
\end{equation}
\begin{equation}
x_{c_{2},2}\left(i\right)=0.5x_{c_{2},1}\left(i\right)+\delta_{x}\left(i\right)
\end{equation}
\begin{equation}
x_{c_{2},3}\left(i\right)=0.2x_{c_{2},1}\left(i\right)+0.4x_{c_{2},2}\left(i\right)+\delta_{x}\left(i\right)
\end{equation}
 {Both signals pass through a linear system with
memory where the output is given by
\begin{equation}
y\left(i\right)=\boldsymbol{r}^{T}\mathbf{x}_{c}\left(i\right)-0.3y\left(i-1\right)+0.35y\left(i-2\right)
\end{equation}
with $\boldsymbol{r}=\left[\begin{array}{ccc} 1 & 0.5 &
0.3\end{array}\right].$ A nonlinear function is then applied to
$y\left(i\right)$
\begin{equation}
d\left(i\right)\begin{cases}
\frac{y\left(i\right)}{3\left(0.1+0.9y^{2}\left(i\right)\right)^{1/2}} & y\left(i\right)\geq0\\
\frac{-y^{2}\left(i\right)\left[1-e^{0.7y\left(i\right)}\right]}{3}
& y\left(i\right)<0
\end{cases}
\end{equation}
The desired signal is corrupted by AWGN with $\sigma_{n}=0.001$.
Fig. \ref{correlated} illustrates the performance of NLR-KNLMS. The
results show that the convergence speed for both inputs is similar.
However, the correlation between the elements of the second input is
stronger than that of the first input. This affects directly the
performance of the NLR-SM-KNLMS as shown in Fig. \ref{correlated},
where we can see that the first input achieves a lower MSE than the
second input.}

 {In the last experiment of this section, we consider
the identification of a Hammerstein system \cite{Greblicki86}. The
input vector ${\mathbf x}(i) \in\mathbb{R}^{1\times24}$ where each
element has $\sigma_{x}^{2}=4\times10^{-4}$ and the noise variance
is $\sigma_{n}=10^{-6}$ . The kernel bandwidth was set to $0.048$.
The input goes through a nonlinear function to form the vector
$\tilde{\mathbf{x}}(i)$, where each element is given by}
\begin{equation}
\tilde{x}_{j}\left(i\right)=x_{j}^{3}\left(i\right)
\end{equation}
 {The desired signal is obtained from a linear system
expressed by}
\begin{equation}
d\left(n\right)=\mathbf{s}^{T}\tilde{\mathbf{x}}\left(i\right)
\end{equation}
 {with $s_{1}=1$, $s_{2}=0.5$, $s_{3}=0.3$,
$s_{4}=s_{5}=s_{9}=s_{13}=s_{15}=s_{19}=s_{22}=0.1$,
$s_{6}=s_{7}=-0.2$, $s_{8}=s_{10}=s_{14}=-0.15$, $s_{18}=0.15$,
$s_{9}=,s_{11}=0.12$, $s_{12}=-0.09$, $s_{16}=0.05$, $s_{17}=-0.05$,
$s_{20}=0.03$, $s_{21}=-0.12$, $s_{23}=-0.02$, $s_{24}=-0.01$.}

\begin{figure}[H]
\begin{centering}
\includegraphics[scale=0.5]{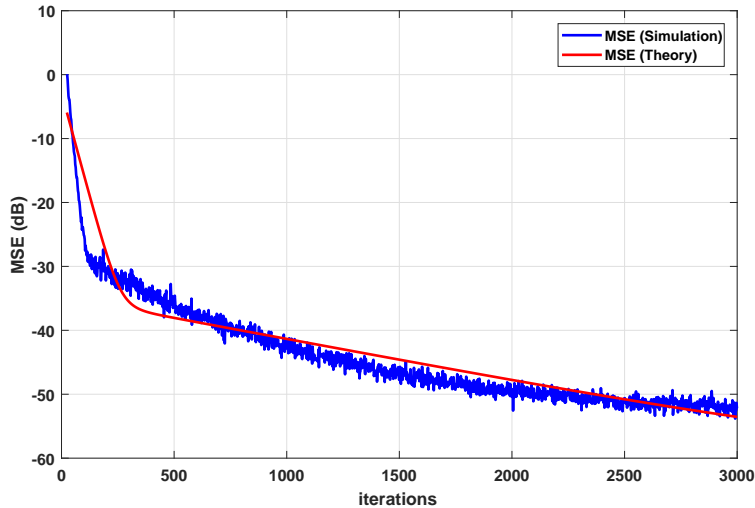}
\par\end{centering}
\caption{Performance of the NLR-SM-KNLMS algorithm for a Hammerstein
system.} \label{Hammer}
\end{figure}

 {The results shown in Fig. \ref{Hammer} indicate that
the learning speed of the proposed NLR-SM-KNLMS is lower than that
for the identification of other nonlinear systems. This is because
the Hammerstein system considered here has a larger number of
parameters, which requires more iteration for the identification.
The curves in Fig. \ref{Hammer} also show that the theoretical
results agree well with those obtained by simulations.}

\subsection{Time series prediction}

Let us now consider the performance of the proposed algorithms for a
time series prediction task. We have used two different time series
to perform the tests, the Mackey Glass time series \cite{Glass79}
and a laser generated time series.  {First, we separate the data
into two sets, one for training and the other for testing as
suggested in \cite{LiuPrincipeHaykin2010}. The time-window was set
to seven and the prediction step was set to one so that the last
seven inputs of the time series were used to predict the value one
step ahead. Additionally, both time series were corrupted by
additive Gaussian noise with zero mean and standard deviation equal
to $0.04$. Using the Silverman rule and after several tests, the
bandwidth of the kernel was optimized and the optimum value found
was one.}

\begin{figure}[H]
\centering{}\includegraphics[scale=0.5]{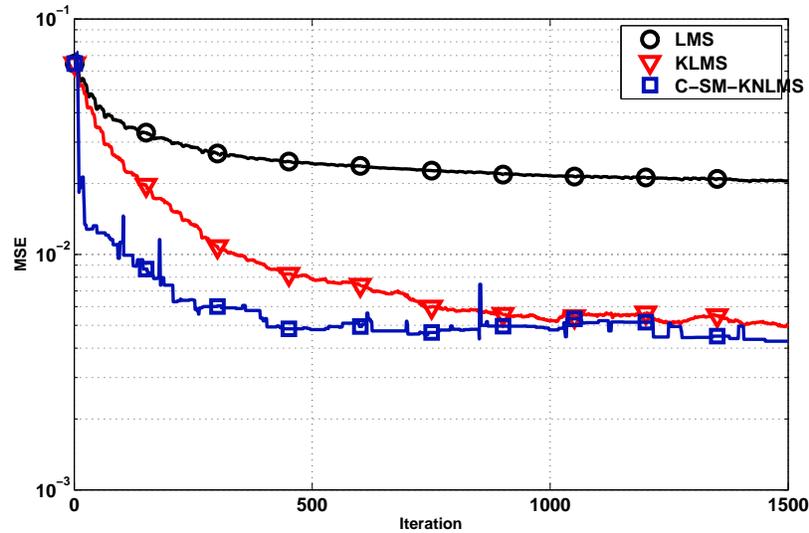}\caption{Learning Curve
of the Kernel Adaptive Algorithms for the Mackey-Glass Time Series
prediction\label{fig:Learning-Curve MG}}
\end{figure}

First we evaluate the performance of the adaptive algorithms over
the Mackey-Glass time series, which is generated by a nonlinear time
difference equation that can be used to model nonlinear dynamics
including chaos and represents a challenging time series for
prediction tasks \cite{Glass79}. A total of 1500 sample inputs were
used to generate the learning curve and the prediction
was performed over 100 test samples. 
For the KLMS algorithm the step size was set to $0.05$. The error
bound for the C-SM-KNLMS algorithm was set to $\sqrt{5}\sigma$. The
final results of the algorithms tested are shown in Table
\ref{tab:Performance MG time series} where the last 100 data points
of each learning curve were averaged to obtain the MSE. The learning
curves of the algorithms based on kernels is presented in Fig.
\ref{fig:Learning-Curve MG}. From the curves, we see that the
proposed C-SM-KNLMS algorithm outperforms conventional algorithms in
convergence speed.

\begin{table}[h]
\caption{Performance on Mackey-Glass time series
prediction\label{tab:Performance MG time series}}
\centering
\begin{centering}
\begin{tabular}{|c|c|c|}
\hline \textbf{Algorithm} & \textbf{Test MSE} & \textbf{Standard}
\textbf{Deviation}\tabularnewline \hline \hline LMS & 0.023 &
+/-0.0002\tabularnewline \hline NLMS & 0.021 &
+/-0.0001\tabularnewline \hline SM-NLMS & 0.020 &
+/-0.0008\tabularnewline \hline
KLMS & 0.007 & +/-0.0003\tabularnewline \hline C-SM-KNLMS & 0.005 &
+/-0.0004\tabularnewline \hline
\end{tabular}
\par\end{centering}

\end{table}

\begin{figure}[H]
\begin{centering}
\includegraphics[scale=0.5]{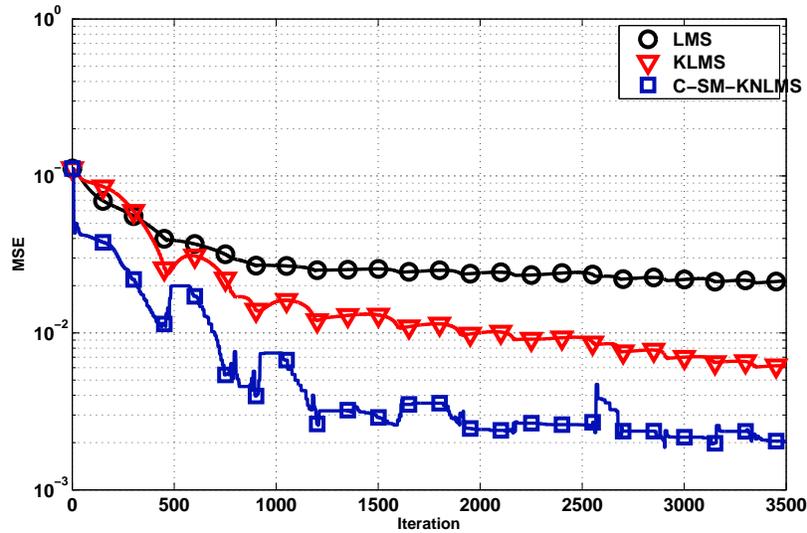}
\par\end{centering}
\caption{Learning curves for the Laser Time Series prediction
\label{fig:Learning-Curve Laser}}
\end{figure}

In the second example of this section, we consider the performance
of the proposed algorithms over a laser generated time series, which
is generated by chaotic intensity pulsations of a laser and also
represents a challenging time series for prediction tasks
\cite{LiuPrincipeHaykin2010}. In this case, $3500$ sample inputs
were used to generate the learning curves and the prediction was
performed over 100 test samples. The setup used in the previous
experiment was considered. Table \ref{tab:Performance Laser time
series} summarizes the MSE obtained for every algorithm tested. The
learning curves are shown in Fig. \ref{fig:Learning-Curve Laser}.

\begin{table}[H]
\begin{centering}
\caption{Performance\label{tab:Performance Laser time series} on
laser generated time series prediction}\vspace{0em}
\end{centering}
\begin{centering}
\begin{tabular}{|c|c|c|}
\hline \textbf{Algorithm} & \textbf{Test MSE} & \textbf{Standard}
\textbf{Deviation}\tabularnewline \hline \hline LMS & 0.021 &
+/-0.0003\tabularnewline \hline NLMS & 0.019 &
+/-0.001\tabularnewline \hline SM-NLMS & 0.024 &
+/-0.006\tabularnewline \hline
KLMS & 0.009 & +/-0.0006\tabularnewline \hline C-SM-KNLMS & 0.003 &
+/-0.0005\tabularnewline
\hline
\end{tabular}
\par\end{centering}

\end{table}

\begin{figure}[H]
\begin{centering}
\includegraphics[scale=0.5]{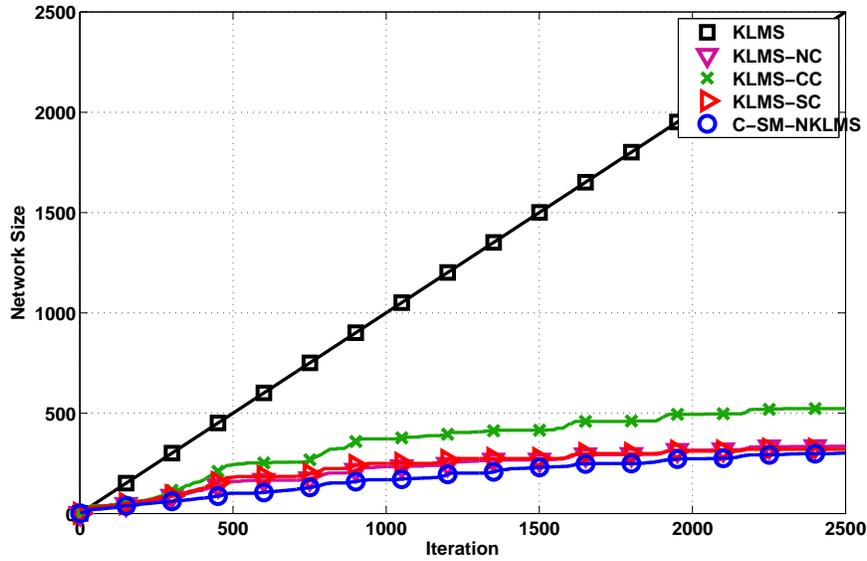}
\par\end{centering}
\caption{Dictionary \label{fig:Dictionary-Size}Size vs Iterations}
\end{figure}

\begin{figure}[H]
\begin{centering}
\includegraphics[scale=0.5]{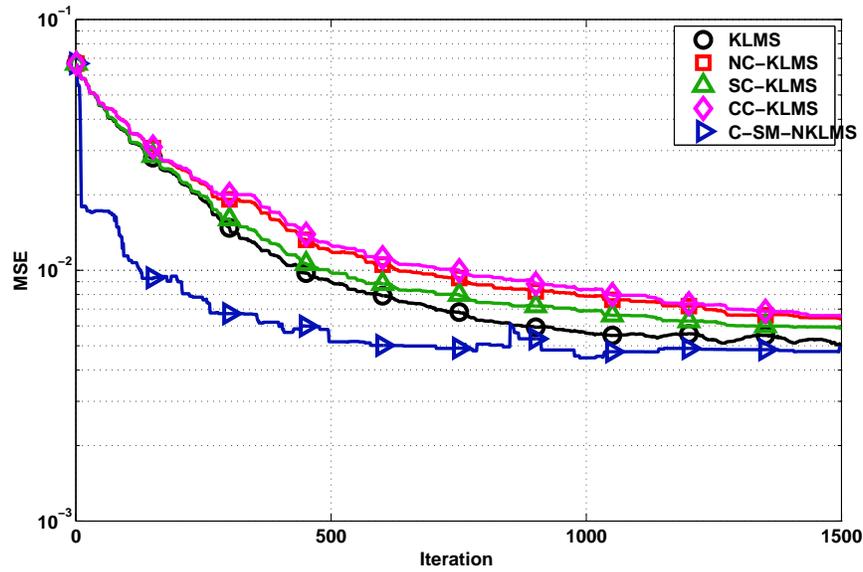}\caption{Performance comparison C-SM-KNLMS vs KLMS over time iterations.
\label{fig:Performance-comparision-SM-KNLMS}}
\par\end{centering}
\end{figure}

In the third experiment of this section we study the size of the
dictionary generated by the conventional KLMS algorithm using
different criteria to limit the size and by the proposed C-SM-KNLMS
algorithm. The result is presented in Fig.
\ref{fig:Dictionary-Size}. We notice that the proposed C-SM-KNLMS
algorithm naturally limits the size of the dictionary. We also
compare the performance of the C-SM-KNLMS with the performance
obtained by the KLMS algorithm with different criteria. Fig.
\ref{fig:Performance-comparision-SM-KNLMS} summarizes the results,
which shows that the proposed C-SM-KNLMS algorithm outperforms the
existing algorithms by a significant margin.

\begin{figure}[H]
\centering{}\includegraphics[scale=0.5]{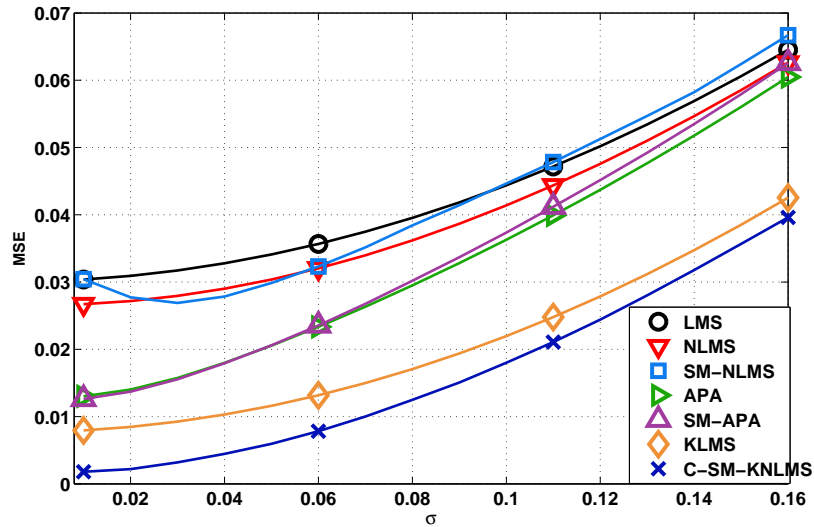}\caption{Robustness
performance of the studied algorithms versus standard deviation of
noise. \label{fig:Robustness}}
\end{figure}

 {In the last experiment, we have assessed the
robustness of the proposed and existing algorithms for Gaussian
noise with different values of standard deviation. Fig.
\ref{fig:Robustness} shows the results in terms of MSE performance
against the noise standard deviation. The curves obtained in Fig.
\ref{fig:Robustness} indicate that the proposed C-SM-KNLMS algorithm
outperforms the other algorithms for all the range of values of
noise standard deviation considered. As expected the performance of
all algorithms evaluated gradually degrade as the noise standard
deviation increases.}

\section{Conclusions}

In this paper, we have devised data-selective kernel-type
algorithms, namely, the centroid-based and the nonlinear regression
SM-KNLMS algorithms. The proposed SM-KNLMS algorithms have a faster
convergence speed and a lower computational cost than the existing
kernel-type algorithms in the same category. The proposed SM-KNLMS
algorithms also have the advantage of naturally limiting the size of
the dictionary created by kernel based algorithms and a satisfactory
noise robustness. {These features allow the proposed SM-KNLMS
algorithms to solve nonlinear filtering and estimation problems with
a large number of parameters without requiring a much longer
training or computational cost. Simulations have shown that the
proposed SM-KNLMS algorithms outperform previously reported
techniques in examples of nonlinear system identification and
prediction of a time series originating from a nonlinear difference
equation.}

\section*{Acknowledgment}

The authors would like to thank the CNPq, and FAPERJ Brazilian agencies for funding.
\\



%

\bibliographystyle{IEEEtran}
\bibliography{adaptivefilters}
\end{document}